\documentclass[]{aa} 
\usepackage{graphicx} 
\usepackage{rotating,subfigure,amssymb,afterpage} 
\usepackage{txfonts} 
\usepackage[pdftitle={XMM Cluster Structure Survey}, 
pdfauthor={H. B\"ohringer et al.}, 
pdfsubject={}, pdfkeywords={}, bookmarks, bookmarksopen,  
colorlinks=true, linkcolor=blue, citecolor=blue, anchorcolor=blue, 
urlcolor=blue]{hyperref}  
\newcommand{\propsim}{\lower 3pt \hbox{$\, \buildrel {\textstyle 
      \propto}\over {\textstyle \sim}\,$}} 
\begin{document} 
   \title{On the Self-Similar Appearance of Galaxy Clusters in X-rays} 
   \author{H. B\"ohringer\inst{1}, K. Dolag\inst{2,3},
   G. Chon\inst{1}} 

%S. Borgani\inst{3} } 

   \offprints{H. B\"ohringer, hxb@mpe.mpg.de} 
 
   \institute{$^1$ Max-Planck-Institut f\"ur extraterrestrische Physik, 
                 D 85748 Garching, Germany, {\tt hxb@mpe.mpg.de}\\ 
              $^2$ Max-Planck-Institut f\"ur Astrophysik,
                 D 85748 Garching, Germany\\
              $^3$ Universit\"atssternwarte M\"unchen, Scheinerstr. 1, 
                 D 81679 M\"unchen, Germany
%             $^4$ Dipartimento di Astronomia dell'Universit\'a di
%               Trieste, via Tiepolo 11, I-34133 Trieste, Italy\\
             } 
 
   \date{Submitted 20/4/11}

\abstract 
{The largest uncertainty for cosmological studies using clusters of galaxies  
is introduced by our limited knowledge of the statistics of galaxy cluster 
structure, and of the scaling relations between observables and 
cluster mass. A large effort is therefore undertaken to compile global galaxy
cluster properties in particular obtained through X-ray observations and to study
their scaling relations. However, the scaling schemes used in the literature differ.}   
{The present paper aims to clarify this situation by providing a thorough review
of the scaling laws within the standard model of large-scale structure growth
and to discus various steps of practical approximations.}
{We derive the scaling laws for X-ray observables and cluster mass within the pure
gravitational structure growth scenario. Using N-body simulations we test the
recent formation approximation used in earlier analytic approaches which involves a
redshift dependent overdensity parameter. We find this approximation less precise than
the use of a fiducial radius based on a fixed overdensity with respect to critical density.} 
{Inspired by the comparison of the predicted scaling relations with observations
we propose a first order
modification of the scaling scheme to include the observed effects of hydrodynamics in 
structure formation. This modification involves a cluster mass dependent gas mass fraction. 
We also discuss the observational results of the reshift evolution of the most important
scaling relations and find that also a redshift dependence of the gas mass to total mass 
relation has to be invoked within our modification scheme.}
{We find that the current observational data are within their uncertainties
consistent with the proposed modified scaling laws.}

\authorrunning{B\"ohringer et al.} 
\titlerunning{Self-Similar Appearance of X-ray Clusters} 
   \maketitle 
% 
%________________________________________________________________ 
 
\section{Introduction} 
 
Galaxy clusters form from overdense regions in the large-scale matter
distribution, which have small amplitudes at early epochs and only
collapsed to objects very recently. In the standard cosmological model the
large-scale matter distribution is described by a random Gaussian field
characterized by a power spectrum with a smoothly changing power law index
over relevant length scales. This implies that the structure evolution will
feature a large degree of self-similarity in scale and time (e.g. Peebles
1980). Consequently galaxy clusters, which form an integral part of this
large-scale structure, also show an imprint of this general self-similarity.
This connection between the framework of the evolution of the gravitating
matter on large scales and galaxy cluster formation and their observed 
structure was realized in early studies, 
e.g. by Gunn \& Gott (1972), Fillmore \& Goldreich
(1984), Bertschinger (1985), Hoffman \& Shaham (1985) and in
simulations e.g. Frenk et al. (1985), Zureck, Quinn \& Solomon 1988,
Efstathiou et al. 1988, West, Dekel \& Oemler 1987). It resulted in a
comprehensive description of the structure of dark matter halos, of which
galaxy clusters are the most massive representatives, in a series of papers
by Navarro, Frenk \& White (1995, 1996, 1997) and follow-up literature. In this 
picture of purely dark matter structure growth, dark matter halos
form a nearly self-similar, two-parameter family, with the two parameters
being mass and a concentration or time-of-formation parameter.
This structural model of clusters describes an average behavior of the
cluster population, where the different statistical realizations of mass 
distributions in the protoclusters produce a significant scatter in the 
observed structural parameters around this mean. Deviations from
the equilibrium state after merger events further contribute to this
scatter.

The addition of baryons to this model leads to a modification of this picture,
which for clusters can be seen as a perturbation of the dark matter structure
evolution. In this sense galaxy clusters mark the
very interesting transition region, where a first order description involving
only the large-scale structure gravitational physics provides a very effective
guideline and the more complicated hydrodynamics, including radiative cooling and
feedback from star formation as well as AGN activity, constitutes 
a perturbative refinement. At smaller scales,
for galaxies the gaseous astrophysics acting on small scales becomes dominant
for the appearance of the visible objects and the observed evolution of
the large-scale structure on galaxy scales becomes very non-linear. It is therefore
on galaxy cluster scales where we can still very successfully apply analytical
descriptions as a useful guideline for the understanding of structure
evolution. The paper builds on this property of galaxy clusters. 

X-ray observations are currently providing the most detailed account of galaxy
cluster structure and are consequently used extensively to test the predictions
of the large-scale structure growth models. However, they do not directly
provide a picture of the dark matter halo distribution, but the distribution
of the hot intracluster medium (ICM) that fills the entire cluster volume
and radiates in X-rays. Therefore the X-ray appearance of clusters includes
aspects of the hydrodynamics how the gas reacts to the dark matter density 
distribution and how the ICM evolves in its thermodynamic properties
(e.g. Voit 2005). One can use
it as a tracer of the dark matter distribution, for example through the assumption
that it is located in the dark matter potentials in hydrostatic equilibrium.
We can thus expect that the X-ray appearance of galaxy clusters is
featuring some hydrodynamic modification compared to the more readily described 
dark matter distribution.  

Because the galaxy cluster formation is so tightly connected to the large-scale
structure evolution and the fact that there is a well described self-similar
structure statistics in the purely gravitational cluster formation model
(Navarro et al. 1997) we can expect that there are simple analytically 
derivable scaling relations for the global X-ray observables as a function of 
cluster mass. These relation have been studied already early by e.g. Kaiser (1986), 
and Evrard \& Henry (1991) and this theoretical work has been supported by simulations
(e.g. Bryan \& Norman 1998, Borgani 2004, Kravtsov et al. 2006, Evrard et al. 2008, 
Stanek et al. 2010, Short et al. 2010, Borgani \& Kravtsov 2010).

With the event of detailed observational studies of cluster structure in X-rays
by means of the advanced X-ray observatories {\sl Chandra} and {\sl XMM-Newton}, 
large sets of observational data on cluster structure and scaling 
relations have become available now and the detailed testing of
the theoretical predictions for the scaling laws is in full swing
(e.g. Markevitch et al. 1998, Arnaud \& Evrard 1999, Mohr \& Evrard 1997, 
Finoguenov et al. 2001, Ikebe et al. 2002, Reiprich \& B\"ohringer 2002, 
Ponman et al. 2003, Ettori et al. 2004, Vikhlinin et al. 2005, Pointecouteau et al. 2005, 
Arnaud et al. 2005, Pratt et al. 2006, Kotov \& Vikhlinin 2006, Zhang et al. 2006, 
Maughan et al. 2006, Maughan 2007, Arnaud et al. 2007, Pratt et al. 2009, 
Mantz et al. 2010, Arnaud et al. 2010, Sun et al. 2011, Reichert et al. 2011). 
An investigation of the relevant literature shows, however, that a number  
different methods are used for the scaling of the data at different redshifts.
The aim of this paper is therefore to critically review these methods
and to determine the best approach based on comparison with simulations
and observations.

In the above mentioned literature mostly analytical formulations of the scaling relations
have been used, based on general considerations of structure formation. In order to provide
the ground for higher precision in the analysis of the evolution of cluster
structure, the simplifications made in the analytical models should be replaced
by tests and calibrations with N-body simulations. This situation can be compared to
that of the theoretical prediction of the dark matter halo (galaxy cluster) mass function,
where the analytical model by Press and Schechter (1974) has paved the way for the general
formulation of the solution for the mass function, but the actual formulae 
now applied, are the result of careful
calibration with N-body simulations (e.g. Jenkins et al. 2001, Evrard et al. 2002,
Warren et al. 2006, Tinker et al. 2008). Here we adopt a similar approach.
We first present the theoretical framework for the description of the evolution
of the scaling relations in the classical form based on the assumption of the
recent formation approximation and compare it to an alternative scheme used in 
the literature. We then resort to the results of N-body simulations to test the 
predictions of the analytical approaches and discuss which of the presently 
used methods in the literature is best. 

In the second part of the paper we compare the theoretical scaling relations to 
observations, inspect the deviations of the observed scaling
relations from the predictions based on dark matter structure evolution (often called
''gravitational scaling relations''), and discuss these deviations in the context
of the influence of hydrodynamical processes. 
We then explore a simply empirical modification scheme of the scaling relation
using a mass dependent depletion factor for the ICM gas to account for the hydrodynamical
scaling effects and compare the so obtained set of scaling relations to observations.
In a last step we consider how this modification should depend on redshift to be
consistent with the observational data. 

The paper is structured as follows. In section 2 we derive the ''gravitational scaling relations''
based on dark matter structure evolution with the assumption that the baryonic matter
follows the dark matter. In section 3 we investigate the dependence of these scaling relations 
on the redshift dependent overdensity parameter in a $\Lambda$CDM cosmology and 
test this model against the method using a fixed overdensity parameter in section 4
using numerical simulations. In section 5 we discuss the redshift dependence of the 
overdensity parameter, which defines the proper fiducial radius of the clusters, 
in the context of numerical studies of the redshift dependence of the concentration 
parameter of galaxy clusters. Section 6 then starts the second part of the paper
where we discuss the modification
of the scaling relations to include hydrodynamical effects. After a comparison
of the description of the evolution of the scaling relations in terms of the 
parameters E(z) and (1+z) in section 7 and a brief comparison with some recent 
simulations in section 8, we provide a comprehensive comparison of scaling relation
results in the literature with the model predictions in section 9.    
Finally section 10 contains a discussion and conclusions. 

%As reference cosmology we use a $\Lambda$CDM model with the parameters: $\Omega_m = 0.268$, 
%$\Omega_{\Lambda} = 0.732$, and $h = 0.704$.

\section{Analytic gravitational self-similar model}

In describing the self-similar structure of galaxy clusters in the purely
gravitational picture, we will first consider the cluster formation by collapse
in an Einstein-deSitter model (EdS), that is a Universe with critical density and zero
cosmological parameter. When we describe the self-similar evolution
for other cosmologies  in the second step, this model is used as reference model.
For illustration, as sketched in Fig. 1, we use a ``top-hat overdensity'' for the
initial conditions characterized by a homogeneous overdensity within a sphere. 
Under realistic conditions protoclusters will have a wide range of morphologies,
but we can reasonably assume, that the morphology distribution with respect to 
the top-hat model is similar for different masses or formation
times. The statistical realizations are strictly
self-similar  only as far as the power spectrum of the density fluctuations, $P(k)$,
is described by a power law.

The theoretical background to the proper scaling has been worked out some
time ago, with a seminal paper being provided for example by 
Kitayama \& Suto (1996). The aspects
of this model, which are crucial for our discussion, are illustrated in Fig. 1,
and will be first interpreted in the frame of an EdS cosmological model. 
The left hand side of Fig. 1 shows a cluster which virializes at redshift zero, where
the point of virialization is in general defined by the time when an ideal homogeneous sphere
would have collapsed to a point. A good time of reference in the collapse process is
the epoch when the overdensity stops expanding and turns around to collapse, which 
happens for this cluster at z=0.78 when the mean matter density inside the protocluster
is about 6.5 times the background density. This is true for all clusters with different mass 
which virialize at z=0. All these clusters finish their formation at the 
same time - twice the turn-around time - and in the same way from the spherical overdensity
with same amplitude to a self-similar structure with same density shape and amplitude
just with a different radial scaling.
Therefore, for clusters formed at the same 
epoch we can defined a fiducial outer radius by a mean cluster density threshold in units
of the mean or critical density of the EdS universe. 

Inspecting now the evolution of 
a cluster that forms at higher redshift, like the cluster sketched on the right hand side of 
Fig. 1, we find a similar picture. The protocluster has a higher mean density when it
forms compared to younger clusters, but the ratio of mean protocluster 
density to the background density at turn-around is 
the same factor of $\sim 6.5$. From the turn-around epoch, cluster collapse will be 
self-similar to the formation at earlier (later) epochs, just scaled to higher (lower)  
density. Also the background density of the universe will evolve self-similarly during cluster 
collapse, since in
the EdS universe we have $\rho_m = \rho_{crit} \propto t^{-2}$. Therefore, if we define
a fiducial outer radius by the radius at which the mean density of the cluster has the 
same overdensity ratio, $\Delta$, to the mean density of the universe, we can compare 
self-similar radii for clusters at different epochs and different sizes. For the
EdS a popular choice for the fiducial radius ,$r_{\Delta}$, is for example the 
virial radius (with $\Delta \sim 18~ \pi^2  \sim 180$; e.g. Peebles 1980).

A self-similar fiducial radius of the clusters is thus defined by:

\begin{equation}
r_{\Delta}^3 = {3 \over 4 \pi \rho_{crit} \Delta} M_{cluster}(r \le r_{\Delta}) 
\end{equation}

%-------------------------------------------------------------
   \begin{figure}
   \begin{center}
   \includegraphics[width=\columnwidth]{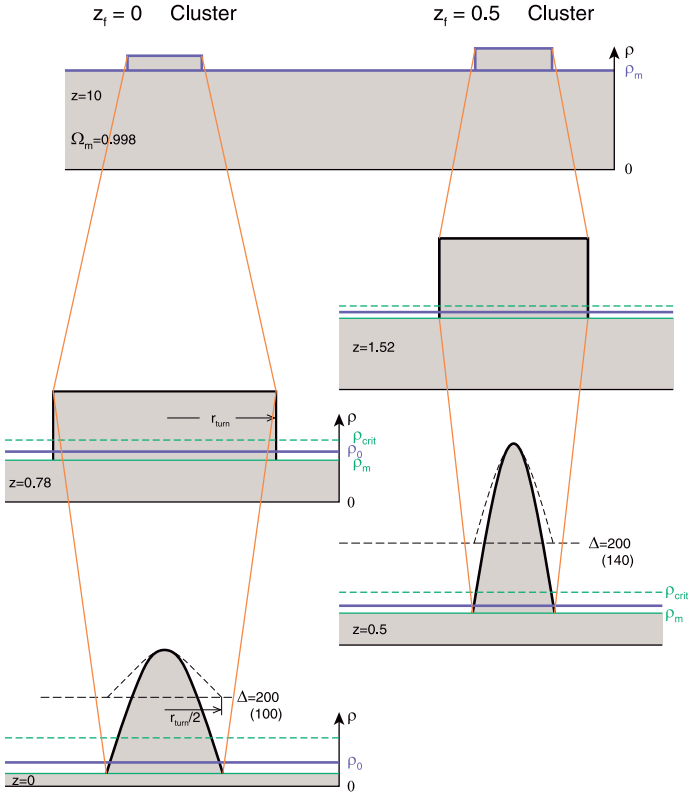}
      \caption{Schematical illustration how the galaxy cluster structure scaling
depends on the formation epoch of the cluster and on the critical density of the Universe 
in an Einstein-de Sitter (densities have blue color) and in a concordance cosmological 
model (densities have green color). The left panel provides a sketch of cluster collapse
for a cluster with a formation redshift of $z=0$ and the right panel shows a 
cluster forming at $z=0.5$. For further explanations see the text.}
         \label{Fig1}
  \end{center}
   \end{figure}
%--------------------------------------------------------------------------------------

The picture becomes more complicated when we change to a low density universe 
(possibly also including a $\Lambda$ term). For the
overdense regions developing into clusters there is no change. According to the Birkhoff
theorem, where a local region of the universe evolves like a universe with these
local density and expansion parameters irrespective of the embedding cosmology,
the cluster evolution does not care about the background universe 
\footnote{The cosmological constant introduces a slight change in the collapse evolution
of the order of a percent, which is neglected here}. Therefore we
can keep the knowledge we have gained in the EdS reference frame, we just have to
introduce another conversion which links the critical density of a general model 
universe with the density of a coevolving EdS universe. To make this more transparent,
we illustrate this point for the case of the 
Concordance Cosmological Model (CCM, with $\Omega_m = 0.3$ and $\Omega_{\Lambda} = 0.7$) 
which seems to provide a close approximation to the structure of the real Universe
(e.g. Spergel et al. 2007, Komatsu et al. 2011).

This is again illustrated in Fig. 1 where a lookback time of 13 Gyr (corresponding
to a redshift of z=10 in the CCM) 
was used as a good approximation to set an EdS and CCM model
to the same density initial conditions and look at their coevolution in time
which is shown in Fig. 2. Note that for a given time the redshifts of the two
models will differ. We see that approaching present time the CCM model has a lower density
than the EdS model, but a higher critical density, which is due to the accelerating 
expansion. Defining the ratios of the critical
densities in Fig. 2 as $\eta(time) = \eta(z_{CCM}) = \rho_{crit}(CCM) / \rho_m(EdS)$, 
where $z_{CCM}$ is the redshift in the CCM (with an index that we will drop 
in the following), we can adapt the overdensity parameter to the new situation by defining

\begin{equation}
\Delta_{CCM}(z) = \Delta_{EdS} /\eta(z)
\end{equation}

and

\begin{equation}
\Delta_{CCM}(z) = \Delta_{CCM}(z=0) * {\eta(z=0)/\eta(z)}
\end{equation}    

Thus, as for example illustrated in Fig. 1, a $\Delta_{EdS} = 200$ in EdS corresponds
to a $\Delta_{CCM} \sim 101$ for $z = 0$ in the CCM model since $\eta(z=0) \sim 2$.
It corresponds to  $\Delta \sim 138$ at $z = 0.5$ in the CCM model, however, since 
with increasing redshift the difference between the two models shrinks. In the following
we drop the index of $\Delta$ and use it only for the CCM.
This formalism
has been worked out in detail for the general case by Kitayama \& Suto (1996),
Eke et al. (1998) and useful approximate formulae for $\Delta(z)$ for a range
of cosmological models including non-flat universes can e.g. be found in Pierpaoli
et al. (2001). 

%-------------------------------------------------------------
   \begin{figure}
   \begin{center}
   \includegraphics[width=\columnwidth]{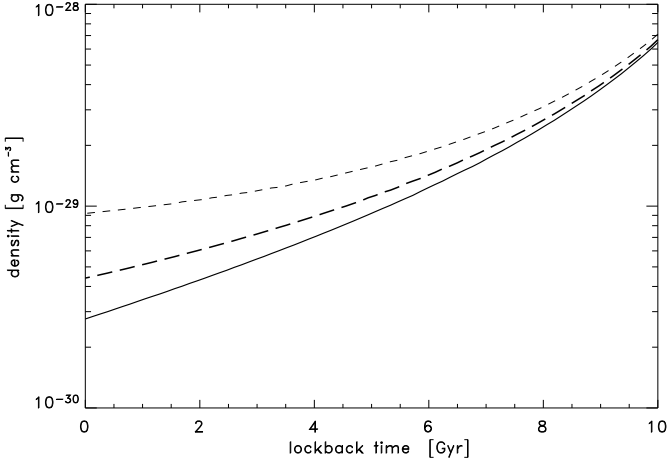}
      \caption{Evolution with lookback time of the mean density, $\rho_m$ and
      critical density $\rho_{crit}$ in the Concordance Cosmological Model
      in comparison with the matter density of a coevolving Einstein-de Sitter 
      reference model, $\rho_0$. The curves are from top to bottom: $\rho_{crit}$,
      $\rho_0$, and $\rho_m$. The two models are compared at the same times after
      the starting point of the coevolving calculations at 13 Gyrs.}
         \label{Fig2}
  \end{center}
   \end{figure}
%--------------------------------------------------------------------------------------

For the dependence of the fiducial radius on mass we have a simple geometrical
scaling, while for the scaling with time we have a proportionality of the mean
density of the cluster to the critical density of the universe, $\rho_0$,
taken either at the time of turn-around or at collapse (for the EdS model).
The mean density as a function of redshift is:

\begin{equation}
{\rho_{critical}(z) \over \rho_{critical}(z=0)}~~ = ~~{H(z)^2 \over H_0^2}~~ \equiv~~ E(z)^2
\end{equation}

The evolution of the radius at fixed overdensity (e.g. $\Delta =200$) is thus given by

\begin{equation}
r_{200}~ \propto~~ \left( M_{200}  \over \rho_0 \right)^{1/3}~~   \propto~~ M_{200}^{1/3} E(z)^{-2/3}
\end{equation}

As mentioned above, the radius at fixed overdensity describes a self-similar region in clusters
at different epochs only in the EdS scenario. Thus if we keep true to the described modeling,
and if we want to compare like with like, we have to use the radii, $r_{\Delta(z)}$
for any comparison. Thus Eq. 5 becomes

\begin{equation}
r_{\Delta(z)}   \propto M_{\Delta(z)}^{1/3} E(z)^{-2/3} {\Delta(z)}^{-1/3}
\end{equation}  

We ought to note here, that in this approach it has been assumed, that the clusters we observe 
have just collapsed. This is true in a very broad sense only, since clusters are always 
accreting material and we always see them in a stage where they have just completed 
some late accretion. This scenario is termed ''recent formation approximation'' in the literature.
A more critical inspection of cluster formation e.g. in N-body simulations shows,
however, that the cluster structure depends in more detail on the whole recent accretion history.
We should therefore be especially concerned about this approximation within the frame of
the presently prefered $\Lambda$CDM cosmology. While in an Einstein-de Sitter cosmological model
structure growth is continuing into the future, it starts to cease in a universe with a
low matter density as soon as the matter density drops below the critical density value.
Thus clusters have accreted matter more slowly in the recent past than at higher redshift
and will practically stop to grow in the distant future (e.g. Busha et al. 2007).  
Thus even though we recognize the beauty and logic of the above approach, we have to critically
test how much deviations are introduced by these approximations by finally comparing
to N-body simulations.

After the radius - mass relation given in Eq. 4, 
the next basic equation is the one linking the X-ray gas temperature with the
cluster mass. The heat of the ICM comes from the conversion of potential
energy during the formation of the cluster. We therefore expect the
temperature to be proportional to the depth of the gravitational potential
and we thus find:

\begin{equation}
T_X~ \propto~ \Phi_0~ \propto~ \left( M_{\Delta(z)} \over r_{\Delta(z)}\right)  \propto
M_{\Delta(z)}^{2/3} E(z)^{2/3} {\Delta(z)}^{1/3}
\end{equation}

In a similar way other essential scaling relations for important X-ray properties can
be constructed (see also Kaiser 1986). In Table 1 we list a set of scaling relations 
involving the cluster mass
or the temperature as scaling parameter (the latter being a prime observable parameter) 
for X-ray determined 
properties as X-ray luminosity, $L_X$ \footnote{For the scaling of 
the X-ray luminosity in a specific band, $L_X$, we assume that the energy band is chosen such,
that the X-ray emissivity is independent of the ICM temperature. This is for example almost fulfilled
for the 0.5 to 2 keV energy band used in most imaging analysis of galaxy clusters, where the change
in emissivity in the temperature range from 2 to 10 keV is less than 6\% for given emission measure.}, 
gas mass, $M_{gas}$, ICM entropy, $K$, density, 
$\rho$, $Y_X$-parameter, with $Y_X = M_{gas} \times T_X$, and pressure, $P$. We have termed
these relations ''gravitational'', since they only include the physics of structure 
evolution of the dark matter which only interacts gravitationally. The baryonic matter
is assumed here to follow the dark matter and hydrodynamical effects have been neglected.
Consequently we will term the relations taking these effects into account as ''hydrodynamical 
relations'', which better describe the observations (e.g. Voit 2005).

In these relations 
the redshift scaling factor $E(z)^2 \times {\Delta(z)}$ appears always in the same
combination just with different powers, and thus one can introduce the abbreviation,
$F(z) \equiv E(z) \times \Delta(z)^{1/2}$. Note that all 
the integral observables or parameters appearing in Table 1, like $L_X$, $M_{tot}$, $M_{gas}$, $Y_X$
have to be integrated out to $R_{\Delta(z)}$, if a comparison between clusters at 
different redshifts are made. Thus, $M$ in the table should actually be written as
$M_{\Delta(z)}$; we have chosen the simplified version of the formulae in
the Table for easier reading.

%_________________________________One column table----------------------------
   \begin{table*}
      \caption{Gravitational scaling relations.}  

         \label{Tempx}
      \[
         \begin{array}{llllll}
            \hline
            \noalign{\smallskip}
 {\rm property~~~~~~~~~~~~}  & {\rm proportionality~~~~~~~~~~~}  & {\rm scaling~ for~ var.} \Delta {~~~~~~~~~~~~~~~~~~~~~~}

        &  {\rm r-dependence^{a}~~}  & {\rm scaling~ for~ fixed}~ \Delta& {\rm deviation}\\
            \noalign{\smallskip}
            \hline
            \noalign{\smallskip}
 {\rm radius,} R_{\Delta}  & \propto M^{1/3} ~F^{-2/3} &  \propto T^{1/2} ~F^{-1}  & \propto r & T^{1/2} ~E^{-1} &    \\
 {\rm density~profile^{a}} & \rho \propto M~ R^{-3}  & \propto F(z)^2  & \propto r^{-2}  & \propto E(z)^2 & -  \\
 {\rm luminosity}^{b}  & L_X \propto \rho^2~R^3~ & \propto T^{1.5} ~F  & \propto r^{1/2} & \propto T_X^{1.5} E~ \Delta(z)^{3/4} & 39\%  \\
 {\rm bolom. lum.}     & L_{bol}  \propto \rho^2~R^3~T^{1/2} & \propto T^2 ~F & \propto r^{1/2} & \propto T_X^{2} E~ \Delta(z)^{3/4} & 39\%  \\
 {\rm entropy,}~ K     & \propto T /\rho^{2/3}   & \propto T ~F^{-4/3} & \propto r & \propto T_X E^{-4/3} \Delta(z)^{-1/6} & 7.5\%  \\
 {\rm gas~mass  }      & \propto \rho~ R^3       & \propto T^{3/2} ~F^{-1} & \propto r & \propto T_X^{3/2} E^{-1}& - \\
   Y_X                 & \propto T~ \rho~ R^3    & \propto T^{5/2} ~F^{-1}   &  \propto r & \propto T_X^{5/2} E^{-1}& - \\
 {\rm pressure }       & \propto  T~ \rho        & \propto  T ~F^2  & \propto r^{-7/3} & \propto  T_X~ E^2 \Delta(z)^{-1/6} & 7.5\% \\
 {\rm surf. bright.}^{a}   & S_X \propto \rho^2~R       & \propto T^{1/2} ~F^{3} & \propto r^{-3} & \propto  T_X^{1/2}~ E^3  & - \\
            \noalign{\smallskip}
            \hline
            \noalign{\smallskip}
 {\rm temperature}     & T_X     & \propto M^{2/3}~  ~F^{2/3}   & (M(r) \propto r)  & \propto M_{\star}^{2/3} E^{2/3}~ & - \\
 {\rm luminosity}^{b}  & L_X     & \propto M  ~F^2~         &   &\propto M_{\star} E^{2}~\Delta(z)^{3/4} & 39\%  \\
 {\rm bolom. lum.}     & L_{bol} & \propto M^{4/3} ~F^{7/3}  &  &\propto M_{\star}^{4/3}~ E^{7/3}~\Delta(z)^{3/4} & 39\%  \\  
 {\rm entropy}         & K       & \propto M^{2/3} ~F^{-2/3} & &\propto M_{\star}^{2/3}~ E^{-2/3}~\Delta(z)^{-1/6} & 7.5\%  \\
 {\rm gas~mass  }      & M_{gas} & \propto M                &  &\propto M_{\star}~ & -   \\
   Y_X                 & Y_X     & \propto M^{5/3} ~F^{2/3} &  &\propto M_{\star}^{5/3}~ E^{2/3}~ & -  \\
 {\rm pressure}        & P       & \propto M^{2/3} ~F^{8/3} &   &\propto M_{\star}^{2/3}~ E^{8/3}~\Delta(z)^{-1/6} & 7.5\%  \\ 
 {\rm surf. bright.}^{a}  & S_X     & \propto M^{1/3} ~F^{10/3} & &\propto M_{\star}^{1/3}~ E^{10/3} & - \\
            \noalign{\smallskip}
            \hline
         \end{array}
      \]
\begin{list}{}{}
\item[]
Columns 1 and 2 give the property and its definition, 
column 3 provides the scaling relation where the parameter $F(z)$ can be read in 
two ways: (i) for the recent formation approximation approach all quantities
which involve a radial integration have to be taken inside $r_{\Delta}$ and
$F(z) \equiv E(z) \times \Delta(z)^{1/2}$ (note that the property of column 1 is for this case 
assumed to be taken at the radius
of the overdensity $\Delta(z)~$), alternatively it can be used for the fixed overdensity model, then all
the radially dependent properties have to be taken at $r_{\Delta}$ and F = E(z).
Column 4 gives the assumed radial dependence of the property of column 1 and column 5 the scaling relation
for the case that the property of column 1 is taken at a fixed overdensity radius. The temperature $T_X$ is in
all cases defined as a mean measured temperature, determined in practical terms usually as a mean temperature in
the region between two defined overdensity radii, except for entropy, $K$ and pressure, where
it means the temperature at the fiducial radius. In general luminosity, gas mass and $Y_X$ are integral
parameters, while density, entropy, pressure, and surface brightness are defined as local parameters at
the fiducial radius. 
$M_{\star}$ is the mass at fixed overdensity. Column 6 lists the deviation of the scaling relation if
the $\Delta(z)$ term is neglected for a comparison of clusters at $z=0$ and $z=1$.
%\begin{list}{}{}
\item[$^{\rm a}$] the columns 4 to 6 refer to the radial dependence of the profile in the radial range $r \sim r_{500}$ to $r_{1000}$  
\item[] Here we assume $\rho \propto r^{-2}$, a $\beta$-model for the gas with $\beta = 2/3$ and $K \propto r$ which 
        implies $T \propto r^{-1/3}$  
\item[$^{\rm b}$] assuming a temperature independent emissivity for the luminosity in a restricted soft X-ray band (see text) 
\end{list}
\label{tab2}
   \end{table*}
%........................................................................................

\section{Dependence on the overdensity parameter}

The appearance of the parameter $\Delta(z)$ in the above equations is a nuisance,
in particular as it can only be calculated by numerical integration or from approximate formulae
given in the literature, e.g. Pierpaoli et al. (2001). Thus some effort has been done 
to check with simulations if this parameter is really necessary, and if the scaling relations
can simply be derived from simulations without a redshift dependent overdensity parameter.
The work by Evrard et al. (2002) and (2008) is for example performed in this spirit.

On the other hand we have seen in the previous chapter, that this parameter has nothing to do
with the cluster formation. It comes solely from the break in the self-similar evolution in 
the background cosmology in going from an EdS to a CCM scenario. Therefore the overdensity
parameter should not be easily abandoned without further checking.

In the following we will investigate if the unwanted parameter can be eliminated
if we make assumptions on the structural parameters of the clusters. 
One of the most fundamental relations is that between temperature 
and mass. Making the very simple assumption that the mass profile is given by an 
isothermal sphere, where $M(r) \propto r$, we find with some arithmetics 
involving Eq. 1 that $M_{\Delta} \propto \Delta^{-1/2}$. 

Eq. 7 can be rewritten as

\begin{equation}
M_{\Delta(z)} \propto T_X^{3/2}~ E(z)^{-1}~ \Delta(z)^{-1/2}
\end{equation}

with the above suggested mass dependence on $\Delta(z)$ we find that

\begin{equation}
M_{\Delta(fix)}(z)= M_{\Delta(z)} \left({\Delta(fix) \over \Delta(z)}\right)^{-1/2}  
\propto T_X^{3/2}~ E(z)^{-1}~ \Delta(fix)^{-1/2}
\end{equation}

where $\Delta(fix)$ is the fixed value for the $\Delta$ parameter independent
of redshift. We note that for the case of the isothermal sphere mass profile the redshift
dependent overdensity parameter is eliminated from the equation of the mass-temperature
relation.

In the more general case of $M(r) \propto r^{\gamma}$ we find

\begin{equation}
r_{\Delta} = r_{200} \left({200 \over \Delta}\right)^{1 \over 3-\gamma}  
\end{equation}

and

\begin{equation}
M_{\Delta} \propto  \Delta^{-\gamma \over 3-\gamma}
\end{equation}

The most widely used model for cluster mass profiles established by simulations
and best confirmed by observations is the NFW model (Navarro et al. 1995, 1997).
\footnote{While recent N-body simulations show deviations from a NFW
profile and promote improvements in form of e.g. an Einasto profile (e.g. Navarro et al.
2004, Gao et al. 2008), the NFW model is sufficiently accurate for our purpose.} 
In Fig. 3 we show the logarithmic slope of this mass profile as a function of radius
and the logarithmic slope for the function $M(\Delta(z))$ for a typical value of
$c = 5$. For the range of interest
for overdensities of 100 to 2500 the slope parameter for the latter function is
in the range 0.3 to 0.6 not far off from the case of the isothermal model and
therefore we can expect an approximate elimination of the  $\Delta(z)$ parameter.
Most of the observational results have been obtained for radii corresponding to
overdensities of 500 for the CCM model which corresponds roughly to an overdensity
of 1000 for EdS at redshift zero and a smaller value for higher redshifts. The
isothermal sphere approximation best applies in the interval 500\ to 1000 where the slope parameter
for the $M(\Delta(z))$ function is in the range 0.4 to 0.5. The maximum deviation
introduced in the comparison between $z = 0$ and $z = 1$ clusters at these
overdensities is of the order
of 4\% and thus far smaller than the uncertainties in all observed relations and
also smaller than the errors in relations derived from simulations. Thus, for the current
precision of the results we can easily neglect the $\Delta(z)$ dependence in the 
mass-temperature relation. This will, however, not be true for any relation e.g. like
those listed in Table 1. In this Table we apply the conversion to $\Delta(fix)$
formulation in a way analogous as done for the mass profile above, with results
shown in column 5. For parameters
like luminosity, entropy and pressure the $\Delta$ dependence does not cancel
for the case of the recent formation approximation. 

The X-ray determined temperature and the velocity dispersion of the galaxies in
optical observations have generally been used as two of the most reliable proxies 
for the estimate of cluster masses \footnote{In the recent literature
notably in Kravtsov et al. 2006 the parameter, $Y_X$, is often promoted as 
the best mass proxy.} (e.g. Arnaud \& Evrard 1999, Carlberg et al. 1996, 
Biviano et al. 2006). Therefore it is interesting that it is exactly for this 
relation, that the $\Delta(z)$ dependence can be neglected. Indeed in the
observational work by Kotov \& Vikhlinin (2006) for example the evolution
of the mass-temperature relation is explained without the need of
$\Delta(z)$, and similarly in the 
simulations by Evrard et al. (2008) they detect a perfect relation for the 
one-dimensional dark matter velocity dispersion and cluster mass of the form

\begin{equation}
M_{200}  \propto  \sigma_v^{2.975 \pm 0.023}~ E(z)^{-1}~~~  .     
\end{equation}

%-------------------------------------------------------------
   \begin{figure}
   \begin{center}
   \includegraphics[width=\columnwidth]{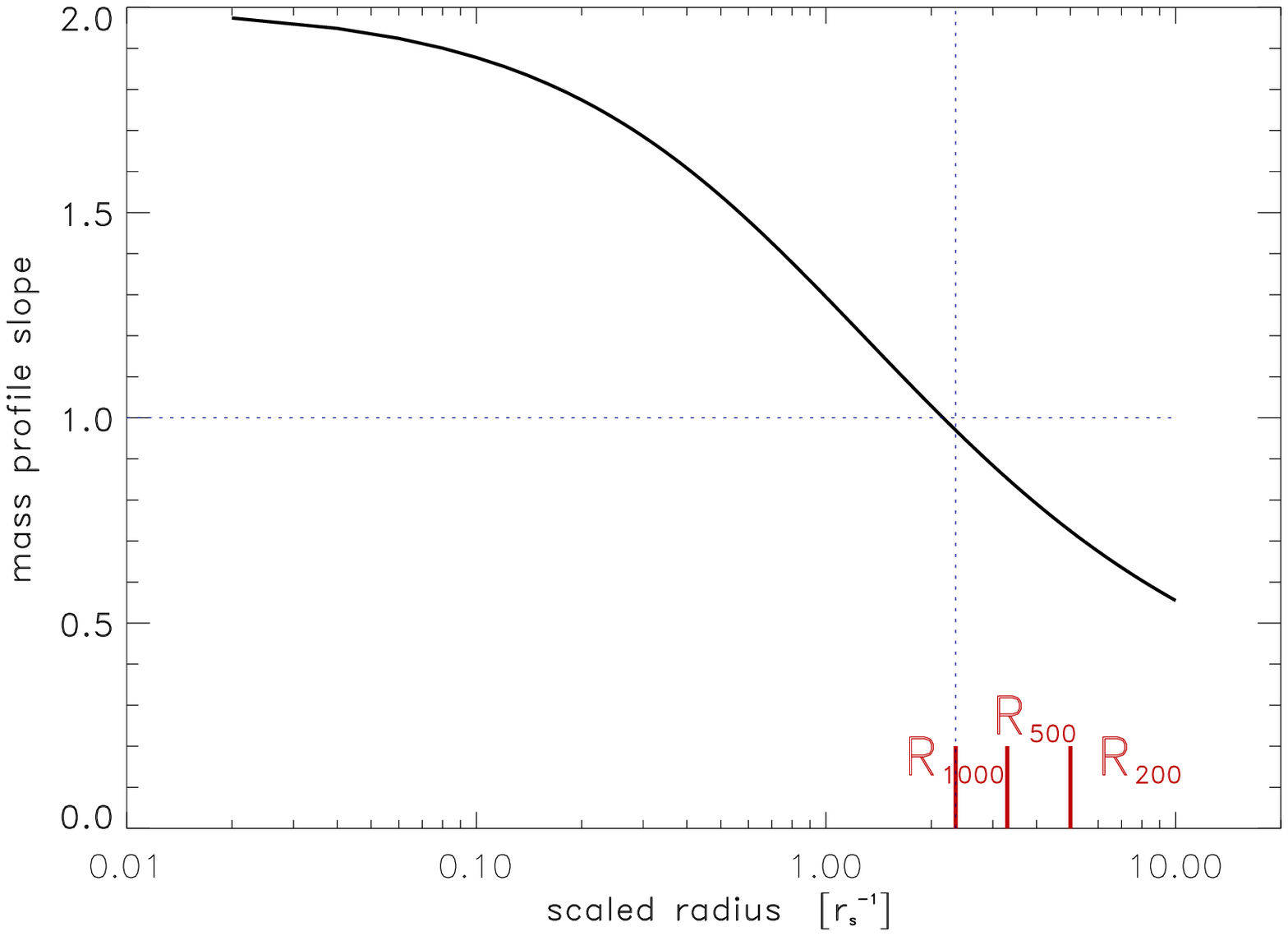}
   \includegraphics[width=\columnwidth]{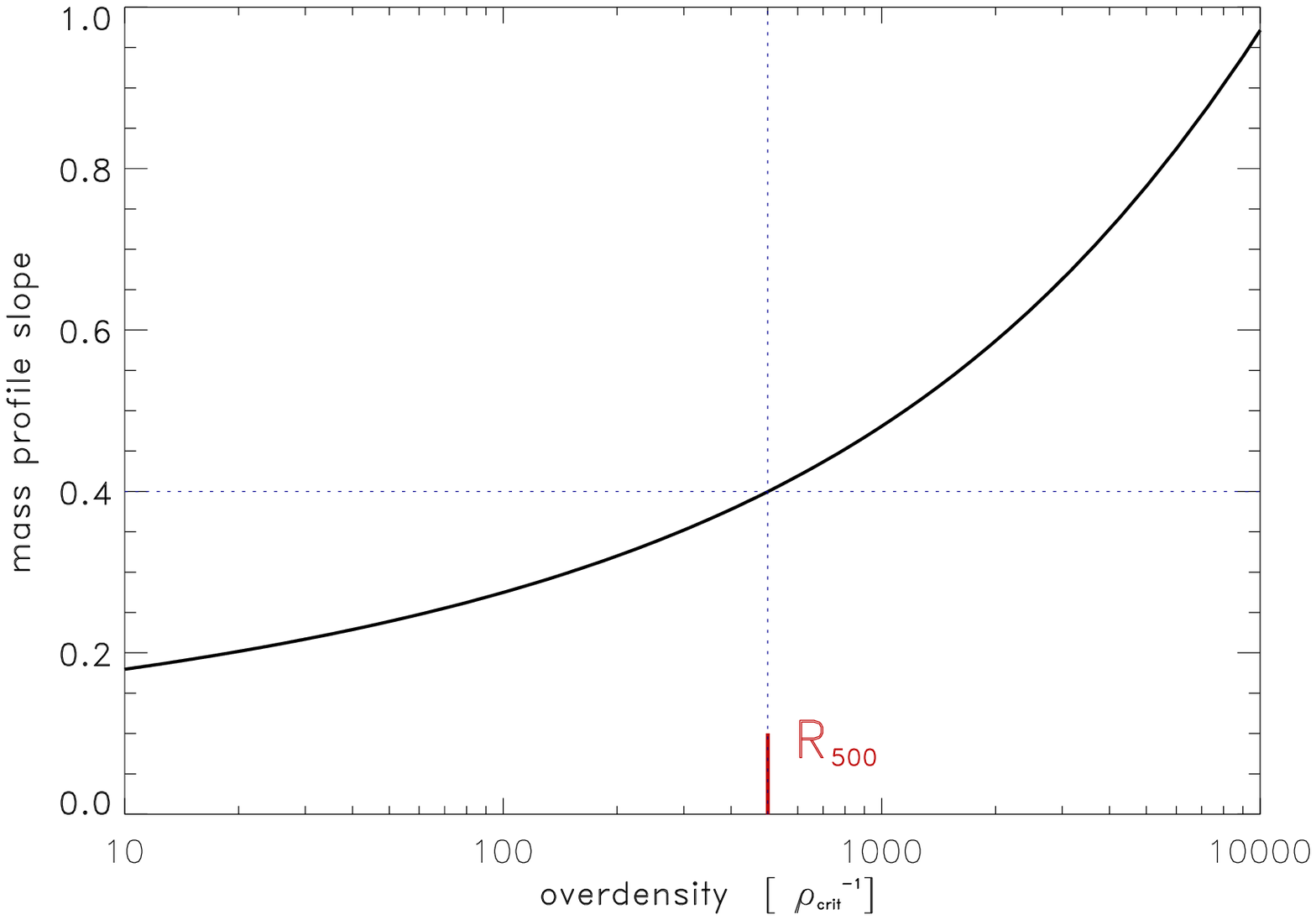}
      \caption{{\bf upper panel:} logarithmic slope of the NFW mass profile as a function
      of scaled radius. {\bf Lower panel:} logarithmic slope of the NFW mass profile a function 
      of the mean density of the cluster over the critical density.}
         \label{Fig3}
  \end{center}
   \end{figure}
%--------------------------------------------------------------------------------------

From these two examples we cannot conclude, however, that $\Delta(z)$ is an unnecessary 
parameter in  general. This will be illustrated through the scaling of density and radius 
with redshift. According to the relations given in Table 1 and from the
illustration in Fig. 1, the density scales as $\rho(z) \propto E(z)^2~ \Delta(z)$.
To see how this works we need a reference density within the density profile of the
cluster. In a cored density profile, we can use the central density, $\rho_0$. For
the NFW density profile, described by 

\begin{equation}
\rho(r) = {\rho_{crit}~ \delta_c \over (r/r_s)~ (1 + r/r_s)^2} 
\end{equation}

where $\rho_{crit}$ is a reference density that is obviously proportional to 
$\Delta(z)$, we can for example use $\rho_s = \rho(r=r_s)$ as reference density
for comparison. This density is characterized by the break of the slope in the 
density profile. If we compare this density point in different clusters it should
have the same overdensity in the EdS cosmology and in the CCM it should scale
according to the above relation. The radius of this point should then scale 
as given by Eq. 6 and expressed as a function of temperature like

\begin{equation}
r_{\Delta(z)} \propto  T^{1/2}~ E^{-1}~  \Delta(z)^{-1/2}
\end{equation}

in which the $\Delta(z)$ dependence can be eliminated according to 
the relations used in Eq. 9 to yield:

\begin{equation}
r_{\Delta(fix)} \propto  T^{1/2}~ E^{-1}
\end{equation}

This then implies that in comparing density profiles we will not observe
a dependence on the  $\Delta(z)$ parameter in the radial scaling
around $r_{500}$ to $r_{1000}$, but we should observe 
the influence of this parameter in the amplitude scaling of the density 
profiles at smaller radii according to Table 1:

\begin{equation}
\rho \propto E(z)^2~ \Delta(z) \propto \rho_{crit}~ \Delta(z)
\end{equation}

From z=0 to z=0.5 (z=1) the parameter  $\Delta(z)$ changes by a factor of
1.36 (1.55). We will test this relation in the next section.

For any observable, $Obs_{\Delta}$, the influence of the  $\Delta(z)$ parameter 
can be investigated in the following way (using a logarithmic Taylor expansion):

\begin{equation}
Obs_{\Delta(z)} = Obs_{\Delta(fix)}~ {\Delta(z) \over \Delta(fix)}
^{{d ln Obs(r) \over d ln r}{d ln r\over d ln \Delta}} 
\end{equation} 

where the last factor in the exponent is again approximately $-0.5$. For example
for the total bolometric luminosity inside $r_{\Delta}$ we find for
a $\beta$-model surface brightness profile with $\beta \sim 2/3$ a behavior of
$L_ {bol}(r)~ \propto~ r^{0.5}$ and consequently

\begin{equation}
L_ {bol}(\Delta(z)) = L_ {bol}(\Delta(fix)~ {\Delta(z) \over \Delta(fix)}
^{-1/4}  
\end{equation} 

For a profile that is steeper than the $\beta$-model, the exponent is even
smaller and the  $\Delta(z)$ dependence is even less important.

In the case of the radial profile for entropy and pressure which depend 
on the temperature profile, we use the observational results that entropy is
approximately proportional to radius (e.g. Pratt 2010), which implies $T_X(r) 
\propto r^{-1/3}$ and the scaling for pressure listed in Table 1. For $Y_X$,
which is defined as gas mass times global temperature and is an integral 
quantitiy, we take the global temperature to be independent of radius which 
then implies that $Y_X$ has the same radial dependence as gas mass and 
total mass.

It is thus clear, that in the frame of high precision cosmology, we cannot
just drop the variable overdensity scenario, as long as the recent formation 
approximation provides a precise picture. Therefore, in the next step, we 
critically test this approximation. 

\section{Testing the density scaling with redshift by N-body simulations}

With the discovery that in the temperature - mass or velocity dispersion - mass relation
the overdensity parameter can be neglected (Evrard et al. 2008), we observe a change
in the literature in the use of scaling relations: in the recent literature
a scaling with fixed overdensity and no overdensity evolution is prefered. Therefore
we will in this chapter test which of these relations is better described by
simulations. We are still concerned with the pure gravitational case and therefore
apply our test to the dark matter distribution. Thus the most fundamental test we can
perform here is to study if the dark matter density distribution scales as 
Eqs. (15) and (16), which we will term the variable overdensity scaling model, 
or just proportional to $E(z)^2$, to which we will refer as fixed overdensity scaling model.

%-------------------------------------------------------------
   \begin{figure}
   \begin{center}
   \includegraphics[width=\columnwidth]{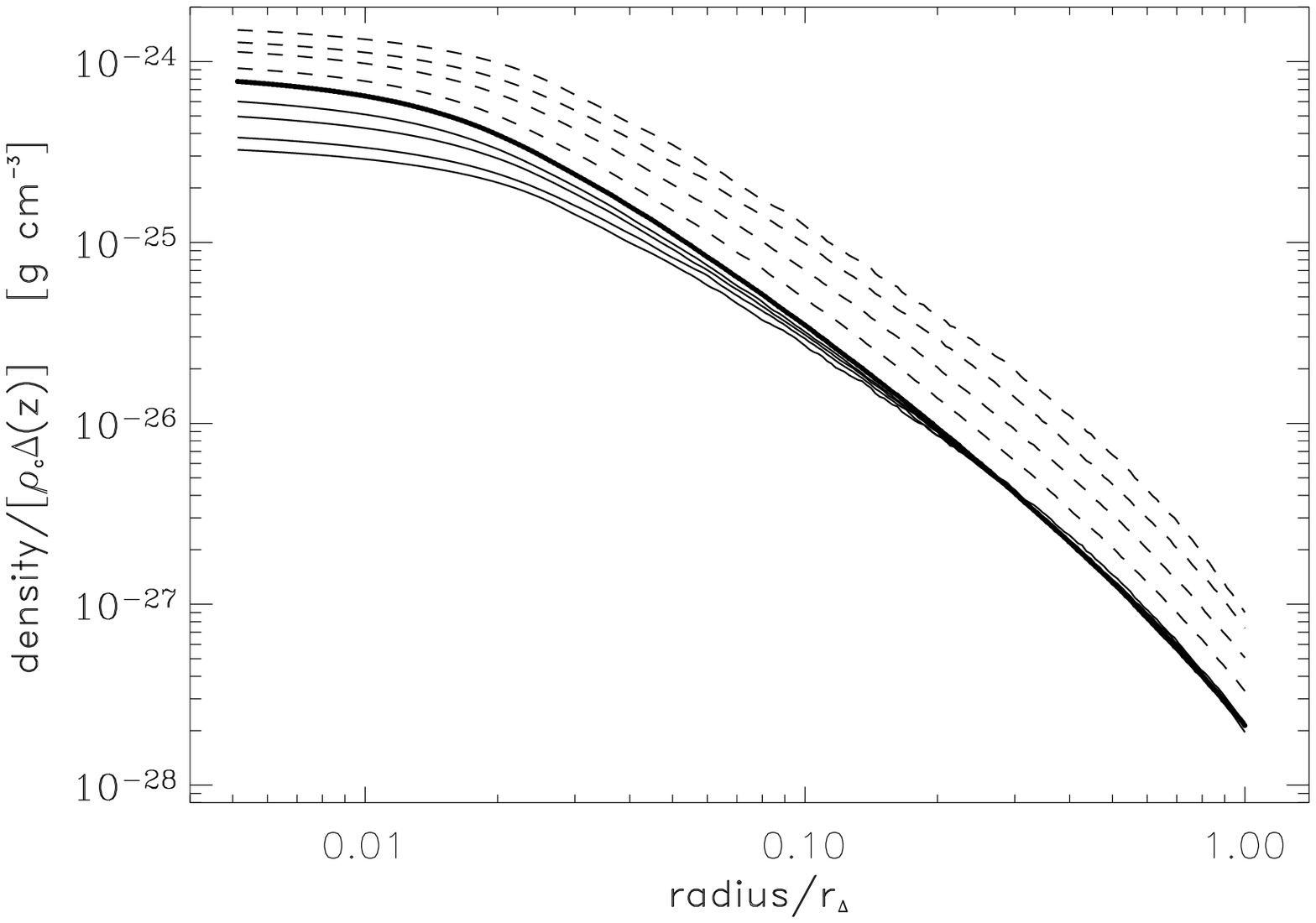}
   \includegraphics[width=\columnwidth]{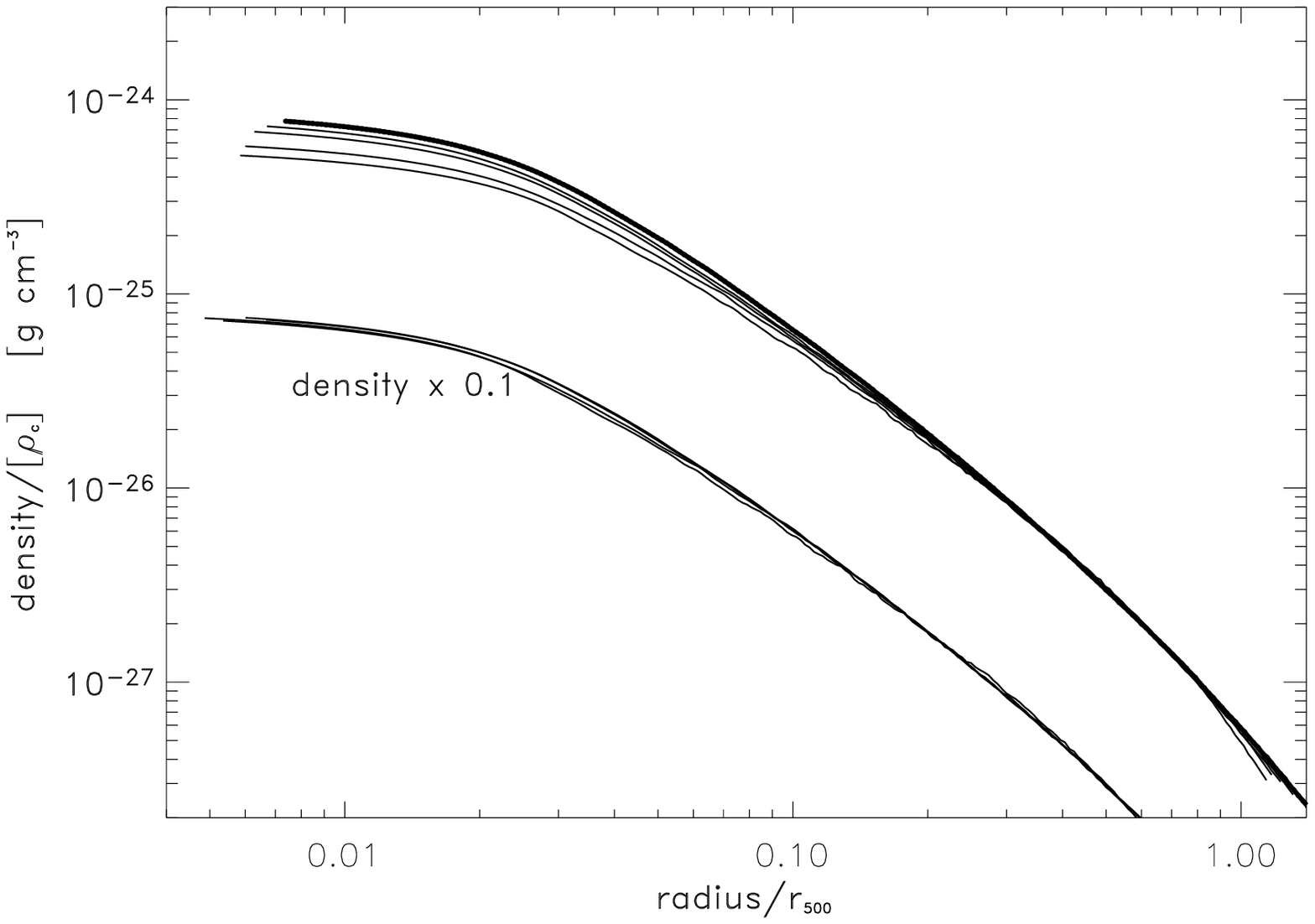}
      \caption{{\bf upper panel:} Testing the redshift evolution of the self-similar scaling of the
    dark matter density distribution with the variable overdensity scaling model.
    The density profiles are given for redshifts of $z = 0$ (heavy line) and
    $z = 0.2521,~ 0.5073,~ 0.7695,~ 1.0013$. Dashed lines show the unscaled and
    solid lines the scaled profiles.  
    {\bf Lower panel:}  redshift evolution of the self-similar scaling of the
    dark matter density distribution with the fixed overdensity scaling model. This model
    clearly works better than the model shown above. We also show, displaced by a factor
    of 0.1, the density profiles after applying the full scaling corrections described
    in section 5. Apart from some numerical fluctuations we observe a perfect fit of
    the scaling corrections.}
         \label{Fig4}
  \end{center}
   \end{figure}
%--------------------------------------------------------------------------------------

For the test we use the simulations by Dolag et al. (2004) which are based on re-simulations
of clusters taken from a large cosmological simulation described in Yoshida et al. (2001)
and Jekins et al. (2001). The cosmological simulation was performed with $512^3$ particles
in a 479 $h^{-1}$ kpc side length box and cosmological parameters of $h = 0.7$, 
$\Omega_{m,0} = 0.3$, $\Omega_{\Lambda} = 0.7$, and a power spectrum normalization of 
$\sigma_8 = 0.9$. The resimulations have a mass resolution 
in the range $2 \times 10^9$ to $6 \times 10^9~ h^{-1}$ M$_{\odot}$ and a gravitational
softening parameter of $5~ h^{-1}$ kpc. The simulation data set used here is for dark matter
particles only to sample the purely gravitational evolution of the clusters.

In Fig. 4 we show the test of the two scaling models. In the upper panel, which shows
the variable overdensity model, we note that the unscaled density increases with increasing
redshift, as expected, but when this increase is corrected by dividing by the factor
$\rho \times \Delta(z)$, we observe an overcorrection of the profiles in the center,
where we can most sensitively test the density scaling. Based on all the reasoning given above,
we have to conclude, that the central density of the clusters does not decrease as much 
with descreasing redshift as expected from the recent formation approximation. More
explicitly, if mass accretion slowly ceases at low redshifts, the clusters keep a more compact
shape than expected in the simplified model. In other words, clusters observed at high redshift
are relatively younger and more recently formed than cluster observed at low redshift.
The recent formation approximation is therefore not a good approximation in a
$\Lambda$CDM model.

The lower panel of Fig. 4 shows the scaling behavior of the fixed overdensity model. It features
much better because the correction made for the density increase with redshift is smaller
and thus the overcorrection is less. We think that there is no more fundamental reason
for the better match of this model than just a fortuitous smaller overcorrection.

At larger radii, $r > 0.15 \times r_{500}$, the profiles rescaled with the variable
overdensity model fit very well and actually slightly better than in the fixed
overdensity case. 

\section{Overdensity scaling based on the evolution of the concentration parameter}    

Since also the fixed overdensity scaling model is not perfectly describing the evolution
of the dark matter density distribution we now seek a more perfect description for the
gravitational scaling relations and take a closer look at the predictions
from N-body simulations. Since a lot of effort was put into an evolutionary description 
of the universal dark matter halo mass profiles, the answer to our problem should be sought
in this approach.

In the literature the change of the shape of the dark matter halo profiles with redshift
is described in the frame of the NFW model (see. Eq. 13) by means of the concentration 
parameter, which depends on the halo mass and the redshift of formation. The
concentration is defined as:

\begin{equation}
c~ =~  {r_{\Delta} \over r_s} 
\end{equation} 
 
where $r_{\Delta}$ is the fiducial cluster radius which can be taken as
$r_{\Delta(z)}$, $r_{200}$, or $r_{m200}$, with $r_{200}$ being the 
radius for a mean overdensity of 200 above critical density
and $r_{m200}$ refers to mean background density (see e.g. Duffy et al. 2008). 
The redshift evolution of this parameter has been studied by Navarro et al. (1997),
Bullock et al. (2001), Eke et al. (2001), Dolag et al. (2004), Duffy et al. (2008),
and Gao et al. (2008). We will use here the results of Dolag et al. (2004)
which comes from the same simulations as used for our testing. The result
they find is well approximated by

\begin{equation}
c_{m200} \propto (1+z)^{-1}
\end{equation}   

The same result is derived by Duffy et al. (2008, Table 1). Gao et al. (2008) do not give the
redshift evolution explicitly, but get similar results with slightly flatter evolution. 
The earlier work finds qualitatively similar results with 
small discrepancies discussed in Dolag et al. (2004). For high precision cosmology
the current exercise should be based on simulations with higher statistics to be
obtained in the future. But the methodic approach will still be the same as outlined here.

We use the findings of Dolag et al. for the concentration parameter to impose
a second order correction to the above relations. We apply the corrections to 
the relations with fixed overdensity, since they have a simpler form and they 
are closer to the simulation results. Thus we have to transform
the behavior of $c_{m200}$ to fixed overdensity scaling with respect to critical 
density, \~c$\equiv c_{200}$.

\begin{equation}
{\rm \tilde{c}}(z) = c_{{m200}(z=0)}~ (1+z)^{-1}~ {r_{200}\over r_{m200}} 
=  c_{m200(z=0)}~ (1+z)^{-1}~ \Omega_m(z)^{1/2}
\end{equation}

Thus we have to correct the radius scaling by the additional factor. 
As clusters get less compact with increasing redshift, their radii will 
be larger than expected and have to be scaled it down accordingly by:

\begin{equation}
r_{scal} = r_{200} \times \left({{\rm \tilde{c}}(z) \over {\rm \tilde{c}}(z=0)}\right)
\equiv r~ \alpha(z)
\end{equation}

To observe mass conservation the change in radial scaling has to be compensated
by a corresponding scaling of the density normalization. For the NFW profile
the density normalization depends on the concentration parameter through the 
proportionality

\begin{equation}
\rho_{c\star} \propto {c^3 \over \left[ln (1+c) - c/(1+c)\right] }
\end{equation}

The amplitude of the density profile will be less than expected with increasing
redshift, as clusters get less compact, and we have to scale the density up
accordingly by:

\begin{equation}
\rho_{scal} = \rho \times \left({\rho_{c\star}(z) \over \rho_{c\star}(z=0) }\right)^{-1}
\equiv \rho~ \beta(z)^{-1}
\end{equation}

%-------------------------------------------------------------
   \begin{figure}
   \begin{center}
   \includegraphics[width=\columnwidth]{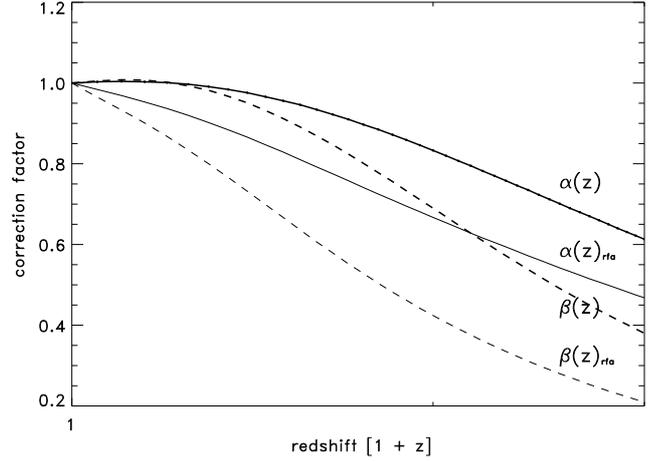}
      \caption{Correction factors $\alpha(z)$ and $\beta(z)$ as a function of redshift
    for the fixed overdensity model (thick lines) and recent formation approximation (rfa)
    model (thin lines and parameters labeled with index rfa).   
    }
         \label{Fig5}
  \end{center}
   \end{figure}
%--------------------------------------------------------------------------------------

The correction terms $\alpha(z)$ and $\beta(z)$ in Eqs. 22 and 24
are shown in Fig. 5 (together with the correction terms
that apply for the recent formation approximation approach).
The magnitude of the correction is 35\% for $\alpha$ and 17\%
for $\beta$ at $z = 1$.
The values are derived for the reference cosmology
model and a mass independent concentration parameter $c(z=0) = 5$.
When these corrections are applied to the density profiles from different
epochs, a perfect scaling within a few percent numerical uncertainty
is obtained, as shown in the lowel panel of Fig. 4.

The correction has a significant effect on the central density, while the effect
is minor at larger radii ($r \ge 0.2 r_{500}$) as can be seen implicitly from 
the lower panel of Fig. 4, as the corrections $\alpha(z)$ and $\beta(z)$ have a
compensating effect. In general the compensating effect for the radial profile
of an  observable can be evaluated (analogous to Eq. 9):

\begin{equation}
Obs_{cor}(z) = Obs(z)~ \beta(z)^{\gamma} ~~ \alpha(z)^{dlog Obs(z) \over dlog r}
\end{equation}

In general we expect this correction to be small (few percent), and we will not
further elaborate on this here, because it will be more useful once we have
a better description of the change of the dark matter halo profiles from
simulations with better statistics.

\section{Modifications due to hydrodynamics}

While the first part of this paper is concerned with the more theoretical aspect, which
model provides the best approach to describe the scaling relations in the purely gravitational
picture of cosmic structure growth, the second part is now exploring how the actually observed
X-ray scaling relations of galaxy clusters can be described within a scenario including
hydrodynamical effects in an empirical way.
Only the most basic relations like those
of ICM temperature or galaxy velocity dispersion with mass
are approximately consistent with the pure gravitational scaling
relation model as given in Table 1. Most other relations involving
ICM properties show deviations. The most
famous of these deviations is the $L_{bol} - T_x$ relation which 
shows an exponent in many observational studies closer to $2.9$
than to the expected value of $2$ (e.g. Edge \& Stewart 1991,
Ebeling et al. 1996, Markevitch 1998, 
Arnaud \& Evrard 1999, Ikebe et al. 2001, Pratt et al. 2009).
A clue to an explanation from the observational side
comes from the fact, that the X-ray surface 
brightness, the line-of-sight integrated emission measure profile,
and the density profile (all very closely related) can be brought to
match surprisingly close at radii $r \ge 0.15 r_{500}$ with an appropriate
amplitude scaling (Arnaud et al. 2002, Croston et al. 
2008). At smaller radii ICM cooling and central AGN feedback is known to 
modify the density profiles (e.g. Fabian 1994, Voit 2005). 
The fact that the shape of the ICM
density profile matches so well, while the scaling of the normalization 
is different from the scaling in Table 1 implies, that the ICM gas mass 
fraction is not constant as a function of cluster mass. An early 
discussion of the change of the gas mass fraction with cluster mass 
can be found in David et al. (1990).

Specifically, Arnaud et al. (2002) find a line-of-sight emission 
measure scaling of the form:

\begin{equation}
EM(r_{scaled})~ \propto~ T^{1.38}~~~~~  {\rm instead~ of:}~~~~~ EM(r_{scaled}) \propto T^{0.5}
\end{equation}

where the second relation stands for the purely gravitational scaling. Since the
integrated emission measure is proportional to the squared density and 
the line-of-sight integration path ($\propto r_{\Delta} \propto T^{0.5}$), 
the amplitude scaling of the ICM density is given by $\propto T^{0.44}$.
Croston et al. (2008) find for the matching of the deprojected density profiles
a best fitting scaling of

\begin{equation}
n_e(r \ge 0.15 r_{500})~ \propto~ T^{0.525}
\end{equation}

for the clusters of the REXCESS sample which span a temperature range
of 2 to 10 keV. Pratt et al. (2009) present directly the gas mass fractions 
of the REXCESS clusters at $r_{500}$ and adding also the results of 
Vikhlinin et al. (2006), Arnaud et al. (2007) and Sun et al. (2009);
they find a best fitting description of

\begin{equation}
f_{gas(500)} \propto M^{0.2}  \propto T^{0.3}
\end{equation} 

The variation of the temperature exponent seen in these results is partly 
an effect of sample variance, but also due to the fact, that the ratio of the
gas mass fraction of clusters of different mass is also a function of radius 
(see e.g. Pratt et al. 2010). 

Neglecting this radial dependence as a higher
order effect, we will explore further the consequences of the mass dependent
gas mass fraction on the scaling relations adopting a mean value of $0.45$ for 
the temperature exponent of the gas fraction relation. Thus we find a
variation of $f_{gas}$ with cluster mass of the form:

\begin{equation}
f_{gas} \equiv {M_{gas} \over M_{tot}}~~ \propto  T^{0.45}~~ 
   \propto M_{tot}^{0.3}
\end{equation}

The decreasing gas mass fraction with decreasing system mass
is explained within the structure formation scenario
by an increasing specific energy introduced into the ICM by star formation and AGN
feedback (e.g. Voit 2005). In more theoretical approaches the reasoning for the modification
of structure is derived from entropy arguments (e.g. Ponman et al. 1999, Bryan \& Voit 2000, 
Voit 2005, Ostriker et al. 2005). For the present derivation given here we prefer the more direct 
observational approach based on the gas mass fraction.

With this observationally implied modification to the scaling relations including
the $f_{gas}$ variation with cluster mass, we can obtain a new set of scaling relations
that are approximately consistent with the observations. These relations are given in 
Table 2 (where column 2 gives the redshift evolution assuming that the $f_{gas} - T_X$
relation does not evolve with redshift).

%_________________________________One column table----------------------------
   \begin{table*}
      \caption{Hydrodynamic scaling relations:} 
         \label{Tempx}
      \[
         \begin{array}{llll}
            \hline
            \noalign{\smallskip}
 {\rm property~~~~~~~~~~~~}  & {\rm proportionality~~~~~~~~~~~}  & {\rm scaling~ no~evoluion~of} f_{gas}~~~~~~~~~ 
        & {\rm scaling~with~empirical~evolution} \\
            \noalign{\smallskip}
            \hline
            \noalign{\smallskip}
 {\rm density~profile}         & \propto \rho ~f_{gas}   & \propto T_X^{0.45} E(z)^2  & \propto T^{0.45} E(z)^{1.385} \\
 {\rm luminosity}^{a}  & L_x \propto \rho^2~R^3~f_{gas}^2 & \propto T_X^{2.4} E           & \propto T_X^{2.4} E^{-0.23} \\ 
 {\rm bolom. lum.}     & L_{bol}  \propto \rho^2~R^3~T_X^{1/2}~f_{gas}^2 & \propto T_X^{2.9} E & \propto T_X^{2.9} E^{-0.23} \\
 {\rm entropy,}~K      & \propto T_X \rho^{-2/3}~f_{gas}^{-2/3}   & \propto T_X^{0.7} E^{-4/3} & \propto T_X^{0.7} E^{1.74}  \\
 {\rm gas~mass  }      & \propto \rho~ R^3 ~f_{gas} & \propto T_X^{1.95} E^{-1}           & \propto T_X^{1.95} E^{-1.615} \\
   Y_X                 & \propto T_X~ \rho~ R^3~f_{gas}  & \propto T_X^{2.95} E^{-1}      & \propto T_X^{2.95} E^{-1.615} \\
 {\rm pressure,}~P(r)  & \propto  T_X~ \rho~f_{gas}    & \propto  T_X^{1.45}~ E^2         & \propto  T_X^{1.45}~ E^{1.385} \\
 {\rm surf. bright.,}~S_X(r) & \propto  \rho^2~R~f_{gas}^2    & \propto  T_X^{1.4}~ E^3 & \propto  T_X^{1.4}~ E^{1.77} \\
            \noalign{\smallskip}
            \hline
            \noalign{\smallskip}
 {\rm temperature}     & T_X     & \propto M^{2/3} E^{2/3}~ &  \\ 
 {\rm gas~mass~fr.}    & f_{gas} & \propto M^{0.3} E^{0.3}~ & \propto M^{0.3} E^{-0.32} \\
 {\rm density~profile}  & \rho & \propto M^{0.3} E(z)^{2.3} & \propto M^{0.3} E^{1.69}  \\ 
 {\rm luminosity}^{a}  & L_X     & \propto M^{1.6}~ E^{2.6}~ &\propto M^{1.6}~ E^{1.37} \\ 
 {\rm bolom. lum.}     & L_{bol} & \propto M^{1.93}~ E^{2.93}~ &\propto M^{1.93}~ E^{2.01}  \\
 {\rm entropy}        & K(r)    & \propto M^{0.47}~ E^{-0.87}~ &\propto M^{0.47}~ E^{0.98}  \\
 {\rm gas~mass  }      & M_{gas} & \propto M^{1.3}~ E^{0.3}~ &\propto M^{1.3}~ E^{-0.32}  \\
   Y_X                 & T_XM_{gas} & \propto M^{1.97}~ E^{0.97}~ &\propto M^{1.97}~ E^{0.35}  \\
 {\rm pressure}        & P(r)    & \propto M^{0.97}~ E^{2.97}~ &\propto M^{0.97}~ E^{2.35}  \\
 {\rm surf. bright.} & S_X(r)  & \propto M^{0.93}~ E^{3.93}~ &\propto M^{0.93}~ E^{2.70} \\
            \noalign{\smallskip}
            \hline
         \end{array}
      \]
\begin{list}{}{}
\item[]
Column 1 and 2 give the property and its definition,
column 3 provides the scaling relation modified for hydrodynamical effects as described in section 6
(note that the property of column 1 is for this case assumed to be taken at the radius 
of the overdensity $\Delta(z)$). For the redshift dependence noted in column 3 the $f_{gas}$ $T_X$ relation
of Eq. (28) is assumed to be redshift independent. For column 4 we then include the redshift dependence
implied from the results of Reichert et al. (2011) as given in Eq. (33).
\item[$^{\rm a}$] for temperature independent emissivity (see Tab.~1 and text)
\end{list}
\label{tab2}
   \end{table*}
%........................................................................................

Compiling all the major scaling relation data from the literature and adding new data for
high redshift clusters from the XDCP project (B\"ohringer et al. 2005, Fassbender 2008)
and from other newly detected distant clusters in Reichert et al. (2011), 
we found first significant constraints on the reshift evolution of some major 
ICM scaling relations. For the mass - temperature relation a result of 

\begin{equation}
M_{500} = 0.291 (\pm 0.031)~ T_X^{1.62 (\pm 0.08)}~ E(z)^{-1.04 (\pm0.07)}
\end{equation}

is found with mass given in units of 
$10^{14}$ M$_{\odot}$ and $T_X$ in units of keV. The relation
shows a redshift evolution that is close to the one expected in the gravitational scenario
(expectation $E(z)^{-1}$). For the relation of bolometric luminosity and temperature
(which is based on two independent observational parameters) they find:

\begin{equation}
L_{bol(500)} = 0.079 (\pm 0.008)~ T_X^{2.70 (\pm0.24)}~ 
E(z)^{-0.23 \left({+0.12 \atop -0.62}\right)}
\end{equation}

and a mass luminosity relation of 

\begin{equation}
  M_{500} = 1.64 (\pm 0.07)~ L_{bol(500)}^{0.52 (\pm 0.03)}~ 
  E(z)^{-0.90 \left({+0.35 \atop -0.15}\right)}
\end{equation}

where $L_{bol}$ is given in units of $10^{44}$ erg s$^{-1}$. The last relation can 
in principle also be derived from the first two, and these results are consistent
with each other within the error limits. 

Therefore we will focus on the implications of the luminosity - temperature relation.
To satisfy the redshift evolution of this relation by means of a redshift dependent
gas mass fraction, we have to imply a relation of the form:

\begin{equation}
f_{gas}~ \propto~ T^{0.45}~ E(z)^{-0.615\left(+0.06 \atop -0.31\right)}
\end{equation}
 
This relation can now be folded into the relations of other properties 
according to their dependence on $f_{gas}$ as listed in Table 2, to find 
their ''empirically'' predicted redshift dependence which is listed in
column 4 of the Table.

\section{Expressions as powers of (1+z)}

%-------------------------------------------------------------
   \begin{figure}
   \begin{center}
   \includegraphics[width=\columnwidth]{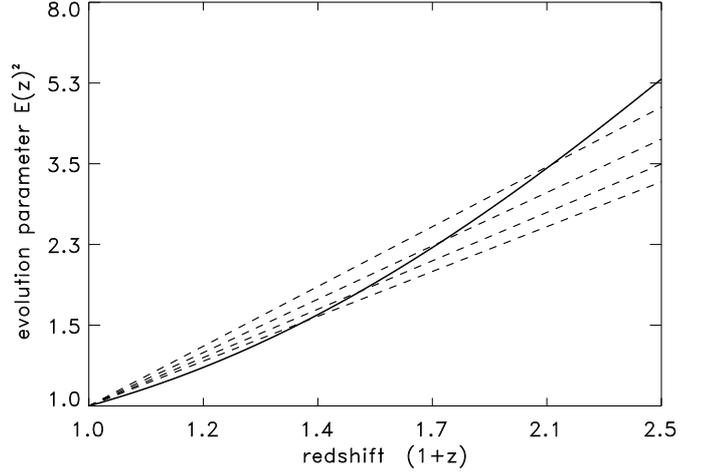}
      \caption{Evolution parameter, 
    $E(z)^2$, as a function of redshift. The dashed  
    lines show the power law functions of (1+z) with exponents
    of 1.26, 1.36, 1.5, 1.68 corresponding to the best approximations
    given in Table 3.}
         \label{Fig6}
  \end{center}
   \end{figure}
%--------------------------------------------------------------------------------------

%_________________________________One column table----------------------------
   \begin{table}
      \caption{Approximations of E(z) in powers of $(1+z)$}
         \label{Tempx}
      \[
         \begin{array}{lll}
            \hline
            \noalign{\smallskip}
{\rm redshift~limit} & {\rm power~of}~E(z) & {\rm max.~deviation}\\
            \noalign{\smallskip}
            \hline
            \noalign{\smallskip}
 0.5   &  0.63   &  0.016 \\
 0.7   &  0.68   &  0.027 \\
 1.0   &  0.75   &  0.045 \\
 1.5   &  0.84  &   0.075 \\ 
            \noalign{\smallskip}
            \hline
         \end{array}
      \]
%\begin{list}{}{}
%\item[$^{\rm a}$] 
%\end{list}
\label{tab3}
   \end{table}
%........................................................................................

In past studies trends of cluster evolution and the evolution of 
scaling relations has most often been modeled with a redshift dependence
of powers $1+z$ in the absence of any better knowledge. For comparison
with these results we studied how well the parameter
$E(z)$ can be expressed by powers
of $1+z$. For the redshift range from zero to a given upper limit we have
determined which power of $1+z$ provides the best approximation with the
smallest maximum deviation from the function. Table 3 provides the results
for upper redshift limits of $z=0.5$, 0.7, 1.0 and 1.5. These approximations and
the functions are shown in Fig. 6. The Figure immediately reveals that the
approximations are not very good, but for limited redshift ranges 
(as typical for the present surveys) the deviations 
are still much smaller than any typical observational uncertainties and 
therefore these simplified relations may still be helpful at the present stage.

Thus for the relation of bolometric luminosity and temperature (in the fixed
overdensity approximation) we find for example:

\begin{equation}
L_{bol} \propto T_X^{2.9}~ (1+z)^{-0.173}
\end{equation}

for the best fit in the redshift range $z = 0 - 1$.

\section{Comparison to some recent simulations}

In the Millenium gas simulations (Springel et al. 2005, Hartly et al. 2008)
the scaling relations were studied by Stanek et al. (2010). For the zero redshift
simulations they find slopes of the scaling relations with mass of
$0.559 (\pm 0.002)$ and $0.576 (\pm 0.002)$ for the mean temperature and spectroscopic
like temperature compared to an expectation value of $2/3$. For $L_{bol}$ and
$Y_x$ they get $1.825 (\pm 0.003)$ and $1.868 (\pm 0.006)$ compared to $2.018$
and $1.9667$, respectively, as given in Table 2. The results are in fair agreement with
the major difference that the $T_X$ - mass relation is less steep in the
simulations which is then also reflected in a slightly shallower slope of all 
other relations.

For the evolution of the parameters they find that $f_{ICM}$, $L_{bol}$ and
$Y_X$ evolve with $E(z)$ with an exponent of $-0.44$, $1.39$, and $0.352$ 
compared to $-0.315$, $1.39$ and $0.33$. Except for $L_{bol}$ there is again
a fair agreement with a slightly larger negative evolution of the gas mass fraction
in the simulations compared to the implication from the results by Reichert et al. (2011).

Using the same basic simulations Short et al. (2010) investigated the dependence of the
scaling relations on the feedback physics used in the simulations. As already discussed
in detail in Reichert et al. (2011), there is a fair agreement of the observations with
the preheating models which involve an early input of energy and elevation of entropy 
of the ICM. In contrast, late feedback models with most of the energy input at redshifts
below 1 are clearly inconsistent with the data. 

\section{Comparison to observations}

In this section we compare the model predictions to recent observations.
We use a representative set of more recently published results and do not aim for a complete
coverage of the literature. In particular the earlier results either suffer from
low statistics or they rely on lower quality observational data. A lot more results on scaling
relations and their evolution with redshift are expected to come in the near future with
a more comprehensive exploitation of the XMM-Newton and Chandra archives and the
completion of several large-survey projects, which will be used for a more critical test
and refinement of the model. 

One of the caveats to keep in mind in the interpretation of the following results
is, that some of the cluster surveys in the literature 
are affected by selection bias effects.
These bias effects arise e.g. from the use of flux-limited surveys, which tend to 
sample preferentially the more luminous clusters in any distribution. We have discussed
and modeled this effect for these types of data sets in the paper by Reichert et al. (2011).
A number of the data sets come from the analysis of galaxy clusters in the data archives;
also these clusters are usually studied as a result of discovery in flux limited surveys.
Only in a few cases e.g. Ikebe et al. (2001) as an example for an earlier paper and
Vikhlinin et al. (2009) have efforts been made to correct for the biasing effects.
In most cases the bias effects on the slope of the relation, which is what concerns us most
here, is smaller than the statistical uncertainties, and therefore we do not consider the
selection bias effects in the following discussion \footnote{In most cases, where the cluster
sample has been compiled from data archives and not from very well defined surveys with published
selection criteria and sensitivity functions, a rigorous reconstruction of the selection effects is not
possible anyway}.

\subsection{Mass - temperature relation}

%_________________________________One column table----------------------------                                  
   \begin{table}
      \caption{Observationally determined slopes of the M - T relation}
         \label{Tempx}
      \[
         \begin{array}{llll}
            \hline
            \noalign{\smallskip}
{\rm relation} & {\rm slope} & {\rm comments}  & {\rm reference}\\
            \noalign{\smallskip}
            \hline
            \noalign{\smallskip}
 M_{500} - T    &  1.64 \pm 0.04    &  {\rm 88~clusters,~ROSAT/ASCA} & {\rm Finoguenov01}\\
 M_{prof} - T   &  1.78 \pm 0.01    &  {\rm                        } & {\rm Finoguenov01}\\
 M_{500} - T^{a)}& 1.48 \pm 0.12    &  {\rm 75~clusters,~ROSAT/ASCA} & {\rm Finoguenov01}\\
 M_{2500} - T   &  1.51 \pm 0.27    &  {\rm 6~relaxed~clusters} & {\rm Allen01}\\
 M_{200} - T^{b)}& 1.84 \pm 0.06    &  {\rm groups~and~clusters} & {\rm Sanderson03}\\
 M_{500} - T    &  1.71 \pm 0.09    &  {\rm 28~clusters,~z = 0.4-1.3} & {\rm Ettori04}\\
 M_{500} - T    &  1.71 \pm 0.09    &  {\rm 10~relaxed~clusters} & {\rm Arnaud05}\\
 M_{500} - T^{c)}& 1.49 \pm 0.15    &  {\rm 6~relaxed~clusters} & {\rm Arnaud05}\\
 M_{200} - T^{d)}&  1.25 \pm 0.37   &  {\rm 11~clusters,~z = 0.6-1.0} & {\rm Maughan06}\\
 M_{2500} - T^{d)}&  2.01 \pm 0.26  &                                 & {\rm Maughan06}\\
 M_{500} - T_m  &  1.58 \pm 0.11    &  {\rm 13~relaxed~clusters} & {\rm Vikhlinin06}\\
 M_{500} - T_{spec}& 1.47 \pm 0.10  &                            & {\rm Vikhlinin06}\\                   
 M_{500} - T    &  1.74 \pm 0.09    &  {\rm 70~clusters,~z=0.18-1.24}  & {\rm O'Hara07} \\ 
 M_{500} - T^{e)}&  1.56 \pm 0.10    &                                  & {\rm O'Hara07} \\ 
 M_{2500} - T   &  1.63 \pm 0.18    &  {\rm 27~RCS~\&~CNOC~clusters} & {\rm Hicks08}\\
 M_{500} - T    &  1.72 \pm 0.18    &  {\rm 13~RCS~clusters, z=0.6-1.1}& {\rm Hicks08}\\
 M_{500} - T^{f)}& 1.65 \pm 0.26    &  {\rm 37~clusters,~XMM-Newton}   & {\rm Zhang08}\\  
 M_{500} - T^{g)}& 1.53 \pm 0.08    &  {\rm 17~clusters,~CHANDRA} & {\rm Vikhlinin09}\\
 M_{500} - T   & 1.76 \pm 0.08      &  {\rm 14~literature~samples}& {\rm Reichert11} \\
           \noalign{\smallskip}
            \hline
         \end{array}
      \]
\begin{list}{}{}                                                                       
\item[$^{\rm a}$ limited to systems with temperature $\ge 3$ keV]
\item[$^{\rm b}$ limited to systems with temperature $\ge 0.6$ keV]
\item[$^{\rm c}$ limited to systems with temperature $\ge 3.5$ keV]
\item[$^{\rm d}$ masses determined assuming isothermal ICM]
\item[$^{\rm e}$ core excised $L_X$ and $T$ with $r\ge 0.2 r_{500}$]
\item[$^{\rm f}$ temperatures for $r = 0.2 - 0.5 \times r_{500}$]
\item[$^{\rm g}$ core excised temperatures with $r = 0.15 - 1 \times r_{500}$]
\item[]

{\tiny
The references are from top to bottom: 
Finoguenov et al. 2001, Allen \& Fabian 2001, Sanderson et al. 2003, Ettori et al. 2004, 
Arnaud et al. 2005, Maughan et al 2006, Vikhlinin et al. 2006, O'Hara et al. 2007, 
Hicks et al. 2008, Zhang et al. 2008, Vikhlinin et al. 2009, Reichert et al. 2011
}                                                                                                                 
\end{list}                                                                                                            
\label{tab4}
   \end{table}
%........................................................................................
                                   
We begin with the comparison for the mass temperature relation. Table 4 provides a large, 
representative, but not complete list of literature results. The results by Reichert et al. 
(2011) are based on a compilation of data from 14 data sets taken from the literature
supplemented by recent published results on individual distant galaxy clusters. It gives 
therefore a summary or average of the largest and most recent observational data samples. 
Most of the values for the correlation slope range from 1.5 to 1.7 for an expected
value of 1.5. The observed relation is thus slightly steeper but it is more close
to the expectation for samples without low mass (low temperature) systems.
For samples with a lower temperature limit above 3 keV the slopes are 
shallower and closer to the self-similar scaling.
 
\subsection{X-ray luminosity - temperature relation}

The luminosity temperature relation is the relation with the two observables 
derived almost independently. The luminosity is obtained from
imaging data with tiny corrections from spectral data, while the temperatures originate
from the interpretation of X-ray spectra. In Table 5 we summarize the observational results
for the luminosity - temperature relation for both, luminosities derived for certain 
energy bands and bolometric luminosities, $L_{bol}$.

%_________________________________One column table----------------------------
   \begin{table}
      \caption{Observationally determined slopes of the L - T relation.} 
         \label{Tempx}
      \[ 
         \begin{array}{llll}
            \hline
            \noalign{\smallskip}
{\rm relation} & {\rm slope} & {\rm comments}  & {\rm reference}\\
            \noalign{\smallskip}
            \hline
            \noalign{\smallskip}
 L_{X1} - T^{a)} &  2.02 \pm 0.4    &  {\rm 35~clusters~ROSAT/ASCA}    & {\rm Markevitch98}\\
 L_{X1} - T^{b)} &  2.10 \pm 0.24   &  {\rm                       }    & {\rm Markevitch98}\\
 L_{bol}- T^{b)} &  2.64 \pm 0.27   &                                  & {\rm Markevitch98}\\
 L_{bol}- T      &  2.88 \pm 0.27   &  {\rm 24~clusters~ROSAT/GINGA}   & {\rm Arnaud99}\\
 L_{X1}- T^{c)}  &  2.47 \pm 0.14   &  {\rm 88~clusters~ROSAT/ASCA}    & {\rm Ikebe02}\\
 L_{bol} - T     &  3.72 \pm 0.47   &  {\rm 28~clusters,~z = 0.4-1.3}  & {\rm Ettori04}\\
 L_{bol} - T    &  2.78 \pm 0.55    &  {\rm 11~clusters,~z = 0.6-1.0}  & {\rm Maughan06}\\
 L_{bol} - T^{f)}&  2.80 \pm 0.2    &  {\rm 115~clusters,~z = 0.1-1.3} & {\rm Maughan07}\\
 L_{bol} - T    &  2.35 \pm 0.33    &  {\rm 70~clusters,~z=0.18-1.24}  & {\rm O'Hara07} \\ 
 L_{bol} - T^{d)}& 2.26 \pm 0.33    &                                  & {\rm O'Hara07} \\ 
 L_{bol} - T    &  2.90 \pm 0.35    &  {\rm 27~RCS~\&~CNOC~clusters}     & {\rm Hicks08}\\ 
 L_{X1} - T^{e)}&  2.13 \pm 0.32    &  {\rm 37~clusters,~XMM-Newton}   & {\rm Zhang08}\\  
 L_{bol} - T^{e)}&  2.61 \pm 0.32   &                                  & {\rm Zhang08}\\  
 L_{X1} - T      &  2.24 \pm 0.22   &  {\rm 31~clusters,~XMM-Newton}   & {\rm Pratt09}\\
 L_{bol} - T     &  2.70 \pm 0.24   &                                  & {\rm Pratt09}\\
 L_{X1} - T^{f)} &  2.32 \pm 0.13   &                                  & {\rm Pratt09}\\
 L_{bol} - T^{f)}&  2.78 \pm 0.13   &                                  & {\rm Pratt09}\\
 L_{bol} - T & 2.53 \pm 0.15        &  {\rm 14~literature~samples}& {\rm Reichert11}\\ 
            \noalign{\smallskip}    
            \hline
         \end{array}
      \]
\begin{list}{}{}
\item[]
$L_{X1}$ is the temperature in the 0.1 to 2.4 keV band, $L_{X2}$ for the 
0.5 to 2 keV band, and $L_{bol}$ is the bolometric luminosity.
\item[$^{\rm a}$ for total luminosity]
\item[$^{\rm b}$ luminosity and temperature corrected for cool core contribution]
\item[$^{\rm c}$ temperature was determined by allowing for an additional] 
\item[cool component not considered in the correlation]
\item[$^{\rm d}$ core excised $L_X$ and $T$ with $r\ge 0.2 r_{500}$]
\item[$^{\rm e}$ temperatures for $r = 0.2 - 0.5 \times r_{500}$]
\item[$^{\rm f}$ core excised $L$ and $T$ with $r = 0.15 - 1 \times r_{500}$]
\item[]
{\tiny References not listed in Table 4 from top to bottom are:
Markevitch 1998, Arnaud \& Evrard 1999, Ikebe et al. 2002, Maughan 2007, 
Pratt et al. 2009}
\end{list}                                   
\label{tab5}
   \end{table}

The band limited luminosity scaling relations have observed slopes in the range $ 2 - 2.5$
with an expected value of about 2.4, while most of the $L_{bol}$ - $T$ relation slopes 
show observed values of $2.6 - 3.7$ with an expected value of about 2.9. The results 
are thus in good agreement with the predictions of the modified scaling relations
within the observational uncertainties.

%........................................................................................

\subsection{Luminosity - mass relation}

The X-ray luminosity - mass relation is one of the most important relations for
cosmological modeling of X-ray cluster surveys. Published results are listed in
Table 6. Again we find good agreement of the observed slopes of 1.4 - 1.7 (for 
$L_{band}$) and 1.6 - 2 (for $L_{bol}$) with 
the predictions for the slope of $1.6$ and $1.93$, respectively.

%_________________________________One column table----------------------------
   \begin{table}
      \caption{Observationally determined slopes of the L - M relation.} 
         \label{Tempx}
      \[ 
         \begin{array}{llll}
            \hline
            \noalign{\smallskip}
{\rm relation} & {\rm slope} & {\rm comments}  & {\rm reference}\\
            \noalign{\smallskip}
            \hline
            \noalign{\smallskip}
 L_{X1} - M_{200}^{a)} & 1.61 \pm 0.09  &  {\rm 106~cluster,~ROSAT/ASCA}   & {\rm Reiprich02}\\
 L_{X1} - M_{200}^{a)} & 1.46 \pm 0.11  &  {\rm 63~cluster,~ROSAT/ASCA}    & {\rm Reiprich02}\\
 L_{bol} - M_{200}^{a)}& 1.84 \pm 0.09  &  {\rm 106~cluster,~ROSAT/ASCA}   & {\rm Reiprich02}\\
 L_{bol} - M_{200}     & 1.90 \pm 0.49  &  {\rm 11~clusters,~z = 0.6-1.0}  & {\rm Maughan06}\\
 L_{bol} - M_{500}     & 1.96 \pm 0.10  &  {\rm 115~clusters,~z = 0.1-1.3} & {\rm Maughan07}\\
 L_{bol} - M_{500}^{b)}& 1.63 \pm 0.08  &                                  & {\rm Maughan07}\\
 L_{X2} - M_{500}^{b)} & 1.45 \pm 0.07  &                                  & {\rm Maughan07}\\
 L_{bol} -  M_{500}    & 1.03 \pm 0.28  &  {\rm 13~RCS~Clusters,z=0.6-1.1} & {\rm Hicks08}\\ 
 L_{bol} - M_{500}     & 2.33 \pm 0.70  &  {\rm 37~clusters,~XMM-Newton}   & {\rm Zhang08}\\  
 L_{X2} - M_{500}      & 1.61 \pm 0.14  &  {\rm 17~cluster,~CHANDRA}       & {\rm Vikhlinin09}\\
 L_{X1} - M_{Y,500}^{c)}& 1.53 \pm 0.10 &  {\rm 31~clusters,~XMM-Newton}   & {\rm Pratt09}\\
 L_{bol} - M_{Y,500}^{c)}& 1.81 \pm 0.10&                                  & {\rm Pratt09}\\
 L_{X1} - M_{Y,500}^{d)} & 1.62 \pm 0.11&                                  & {\rm Pratt09}\\
 L_{bol} - M_{Y,500}^{d)}& 1.90 \pm 0.11&                                  & {\rm Pratt09}\\
 L_{X1} - M_{500}      & 1.64 \pm 0.12  &  {\rm 31~clusters,~XMM-Newton}   & {\rm Arnaud10}\\
 L_{X1} - M_{500}^{d)} & 1.76 \pm 0.13  &                                  & {\rm Arnaud10}\\
 L_{bol} - M_{500}     & 1.51 \pm 0.09  &  {\rm 14~literature~samples}     & {\rm Reichert11}\\ 
            \noalign{\smallskip}    
            \hline
         \end{array}
      \] 
\begin{list}{}{}
\item[]
$L_{X1}$ is the temperature in the 0.1 to 2.4 keV band, $L_{X2}$ for the 0.5 
to 2 keV band, and $L_{bol}$ is the bolometric luminosity.                                                               
\item[$^{\rm a}$ Quoted is the BCES orthogonal fit.]
\item[$^{\rm b}$ Core excised luminosity, $r = 0.15 - 1 \times r_{500}$]
\item[$^{\rm c}$ $M_Y$ is the mass estimated from the $Y_X$ - M relation.]
\item[$^{\rm d}$ Corrected for Malmquist bias]
\item[]
{\tiny References not listed in Tables 4 and 5 from are:
Reiprich \& B\"ohringer 2002, Arnaud et al. 2010}
\end{list}                                        
\label{tab6}
   \end{table}
%........................................................................................

\subsection{Entropy - temperature relation}

Table 7 lists the results for the entropy temperature relation from the observational
analysis of Pratt et al. (2010). We note that the slope of the relation depends on 
the scaled radius at which the measurement is taken. In contrast to the observational
parameters above which are integrated values like e.g. luminosity, this is the local
entropy value at a given radius. The change in slope originates from the fact that
the gas density reduction in low mass systems is not completely self-similar with radius, 
somewhat contrary to what is suggested e.g. by the very tight scaling of the density 
profiles in Fig. 2 of Croston et al. (2008).  

The gravitational scaling model would predict the relation $K \propto T$, while 
our modified scaling relation predicts $K \propto T^{0.7}$. The latter prediction
is approximately met at intermediate radii. For larger radii the results seem
to approach the gravitational scaling relations.

%_________________________________One column table----------------------------
   \begin{table}
      \caption{Observationally determined slopes of the Entropy - T and entropy - mass relation}
         \label{Tempx}
      \[ 
         \begin{array}{llll}
            \hline
            \noalign{\smallskip}
{\rm relation} & {\rm slope} & {\rm comments}  & {\rm reference}\\
            \noalign{\smallskip}
            \hline
            \noalign{\smallskip}
 K_{500} - T   &  0.92 \pm 0.24      & {\rm 31~clusters~XMM-Newton} & {\rm Pratt10}\\ 
 K_{1000} - T   &  0.83 \pm 0.06     &                               & {\rm Pratt10}\\
 K_{2500} - T   &  0.76 \pm 0.06     &                               & {\rm Pratt10}\\ 
 K_{1000} - T   &  0.83 \pm 0.06     &                               & {\rm Pratt10}\\
 K_{500} - M_{500}&  0.62 \pm 0.17   &                               & {\rm Pratt10}\\ 
 K_{2500} - M_{2500}&  0.42 \pm 0.05 &                               & {\rm Pratt10}\\ 
            \noalign{\smallskip}    
            \hline
         \end{array}
      \] 
\begin{list}{}{}                                         
\item[$^{\rm a}$ limited to systems with mass $\ge 3 \times 10^{13}$ M$_{\odot}$]
\item[$^{\rm b}$ limited to systems with temperature $\ge 0.6$ keV]
\item[]
{\tiny The references is Pratt et al. 2010}
\end{list}                                   
\label{tab6}
   \end{table}
%........................................................................................

\subsection{Gas mass - temperature relation}

Some results for the gas mass - temperature relation are listed in Table 8.
The observed range of values for the slope of 1.88 - 2.22 corresponds well to the
predicted value of 1.95.

%_________________________________One column table----------------------------
   \begin{table}
      \caption{Observationally determined slopes of the $M_{gas}$ - T relation}
         \label{Tempx}
      \[ 
         \begin{array}{llll}
            \hline
            \noalign{\smallskip}
{\rm relation} & {\rm slope} & {\rm comments}  & {\rm reference}\\
            \noalign{\smallskip}
            \hline
            \noalign{\smallskip}

 M_{g, 200} - T  &  2.22 \pm 0.31  &  {\rm 11~clusters,~z = 0.6-1.0} & {\rm Maughan06}\\
 M_{g, 2500} - T &  1.80 \pm 0.49  &                                 & {\rm Maughan06}\\
 M_{g, 500} - T  &  2.12 \pm 0.12  &  {\rm 31~low~z~cl.,~XMM-Newton} & {\rm Croston08}\\
 M_{g, 500} - T^{a)}& 1.86 \pm 0.19&  {\rm 37~clusters,~XMM-Newton}   & {\rm Zhang08}\\  
            \noalign{\smallskip}    
            \hline
         \end{array}
      \] 
\begin{list}{}{}                                         
\item[$^{\rm a}$ Temperature for $r = 0.2 - 0.5 \times r_{500}$]
\item[]
{\tiny The reference not listed in previous Tables is Croston et al. 2008}
\end{list}                                   
\label{tab6}
   \end{table}
%........................................................................................

\subsection{Other relations}

Observational results for some other interesting relations are presented in
Table 9. The $Y_X$ parameter is a quantity motivated by cluster observations in
the Sunyaev-Zeldovich effect (SZE). Since the SZE signal is proportional to
the intracluster plasma temperature and the total number of electrons, the 
$Y_X$-parameter is defined as $Y_X = T \times M_{gas}$. This parameter has been
promoted as a very good mass proxy for galaxy clusters (Motl et al. 2005,
Kravtsov et al. 2006).

Values for the slope of the $L_{bol}$ - $Y_X$ relation of 0.65 - 1.1 and
for the $L_{band} - Y_X$ relation of 0.8 - 1.14 agree well with the 
predictions (combining the $L_X - T$ and $Y_X - T$ relations) of values 
of 0.98 and 0.81, respectively. The $M_{500} - Y_X$ relation with
slope values of 0.57 - 0.62 is also not far off from the precdited value of 0.51.

Finally the $M_{tot} - M_{gas}$ relation with a predicted slope of 0.76
is shallower than the listed result of 0.9 by Zhang et al. (2008) and also
somewhat shallower than the slope inferred from Fig. 9 in Vikhlinin et al. (2009)
which roughly corresponds to a slope of 0.83 in the mass range $10^{14}$ to
$10^{15}$ M$_{\odot}$. But our prediction is very well supported by (and partly
originates from) the results of Pratt et al. (2009) based on a large and
representative sample of galaxy clusters with high quality XMM-Newton data.

%_________________________________One column table----------------------------
   \begin{table}
      \caption{Observationally determined slopes for various relations}
         \label{Tempx}
      \[ 
         \begin{array}{llll}
            \hline
            \noalign{\smallskip}
{\rm relation} & {\rm slope} & {\rm comments}  & {\rm reference}\\
            \noalign{\smallskip}
            \hline
            \noalign{\smallskip}
 L_{bol} - Y_X     & 1.10 \pm 0.04  &  {\rm 115~clusters,~z=0.1-1.3} & {\rm Maughan07}\\
 L_{bol} - Y_X^{a)}& 0.94 \pm 0.03  &                                  & {\rm Maughan07}\\
 L_{bol} - Y_X     & 0.65 \pm 0.10  &  {\rm 13~RCS~clusters,~z=0.6-1.1}& {\rm Hicks08}\\
 L_{bol} - Y_X     & 0.95 \pm 0.08   &  {\rm 37~clusters,~XMM-Newton}   & {\rm Zhang08}\\
 M_{500} - M_{gas} & 0.906 \pm 0.08  &                                  & {\rm Zhang08}\\  
 M_{500} - Y_X     & 0.62  \pm 0.06  &                                  & {\rm Zhang08}\\
 M_{500} - Y_X     & 0.57 \pm 0.03    &  {\rm 17~clusters,~CHANDRA}    & {\rm Vikhlinin09}\\
 L_{X1} - Y_X      & 0.84 \pm 0.05   &  {\rm 31~clusters,~XMM-Newton}   & {\rm Pratt09}\\
 L_{bol} - Y_X     & 0.99 \pm 0.05   &                                  & {\rm Pratt09}\\
 L_{X1} - Y_X^{b)} & 0.82 \pm 0.03   &                                  & {\rm Pratt09}\\
 L_{bol} - Y_X^{b)}& 0.97 \pm 0.03   &                                  & {\rm Pratt09}\\
 L_{X1}   - Y_X     & 1.14 \pm 0.08   &  {\rm 31~clusters,~XMM-Newton}   & {\rm Arnaud10}\\
 L_{X1}   - Y_X^{c)}& 1.07 \pm 0.08   &                                  & {\rm Arnaud10}\\
            \noalign{\smallskip}
            \hline
         \end{array}
      \] 
\begin{list}{}{}
\item[$^{\rm a}$ Core excised luminosity, $r = 0.15 - 1 \times r_{500}$]
\item[$^{\rm b}$ core excised $L$ and $T$ with $r = 0.15 - 1 \times r_{500}$]
\item[$^{\rm c}$ Corrected for Malmquist bias]
\end{list}
\label{tab6}
   \end{table}
%........................................................................................

\subsection{Evolution with redshift}

Vikhlinin et al. (2009) quote a result for the evolution of the $L_{X2} - M$ 
relation of

\begin{equation}
L_{X2}~ \propto~ M^{1.61\pm 0.14}~ E(z)^{1.85\pm 0.42}
\end{equation}

which has to be compared to Eq. (32) based on the result of Reichert et al. (2011).
Inverting relation (32) and converting from $L_{bol}$ to $L_{X2}$ we find:
$L_{X2} ~\propto ~L_{bol} T^{-0.5} ~\propto ~T^{-0.5} M^{1.92} E(z)^{1.73}$.
Substituting Eq. (7) without the $\Delta$-factor we get:

\begin{equation}
L_{X2}~ \propto~ M^{1.59\left(+0.11 \atop -0.09\right)}~
 E(z)^{1.40\left(+0.3 \atop -0.7\right)}
\end{equation}

showing agreement between the two results within the uncertainties.

More effort has been put on the study of the evolution of the luminosity - temperature relation.
Here we find a different picture, such that most of previous works finds a positive evolution.
In the past literature, the evolution has most often been parameterized in the form:

\begin{equation}
L_X ~ \propto~  T^{\beta}~~ (1+z)^{\alpha}~~ .
\end{equation}

Taking this functional form and the correspondence of $E(z) \sim (1+z)^{0.75}$ from Table 3 for
the redshift range $z = 0 - 1$ we find for the results of Reichert et al. (2011) a proportionality
of the evolution of the $L_X - T$ relation of $\propto~ (1+z)^{-0.17\left(+0.09 \atop -0.46 \right)}$.
Earlier works find $\alpha = 1.5 \pm 0.3$ in Vikhlinin et al. (2002) and Lumb et al. (2004), and
$\alpha = 1.8 \pm 0.3$ in Kotov $\&$ Vikhlinin (2005). Maughan et al. (2006) find a value of
$\alpha = 1.3 \pm 0.2$ if they combine their WARPS cluster sample with that of Vikhlinin et al. (2002)
and a value of $\alpha = 0.8 \pm 0.4$ for the WARPS sample alone. Ettori et al. (2004) find the
least positive evolution with values for the exponent of (1+z) in the range 0.04 to 0.98
when using Markevitch (1998) and Arnaud $\&$ Evrard (1999) as local reference, but partly
negative values in the range -0.48 to +0.54 when combining their results with those of
Novicki et al. (2002). Common to all these studies
is their use of the data of Markevitch (1998) and Arnaud $\&$ Evrard (1999) for the local ($z \sim 0)$
reference. The normalization of the $L_X - T$ relation in the latter two works is lower by
a factor of very roughly 1.5 than that of Pratt et al. (2009) which has been used in Reichert
et al. as the most important local reference. Therefore, the most obvious reason for the fact
that these earlier works find a positive evolution in contrast to the slightly negative evolution
found by Reichert et al. (2011) is the different local reference in addition
to the small sample sizes and selection bias effects.

Pacaud et al. (2007) also quote an $L_X - T$ evolution best fit by $\alpha = 1.5 \pm 0.4$, but
the fit is poor and the data are different from this approximation at $z \ge 0.7$ as can be
seen in their Fig. 4. Maughan et al. (2011) in a recent paper also find
a positive evolution with substantial deviations from the overall trend at 
intermediate redshifts. 
Different from these other results O'Hara et al. (2007) find a
negative evolution with a value of $\alpha = -0.25 \pm 0.56$.

In summary, it is clear that the sparsity and inhomogeneity of the data used in the past to
study the redshift evolution of the scaling relations lead to inconclusive results and we are
just beginning to see some trends now.  The uncertainties of
the parameterized evolution of the relations are still very large and e.g. in the case of the
$L_X - T$ relation the results are still consistent with no evolution of the $L_X - T$
relation.

\section{Discussion and Conclusion}

Studying in detail the evolution of the dark matter mass density profiles of simulated
galaxy clusters, we have shown that the older model of self-similar scaling relations based on
the recent formation approximation which uses a scenario where the fiducial overdensity radius,
$r_{\Delta(z)}$, is taken to be redshift dependent, is not accurate and a scaling with a fixed
overdensity provides a better and currently sufficiently precise description of self-similar
evolution. For a precise analysis of future cosmological surveys of the galaxy cluster population,
we should improve this description further. We plan to do this with larger N-body/hydrodynamical 
simulations which are performed at present and therefore give no detailed recipes for the 
scaling corrections in this paper. The corrections as shown in Fig. 5 depend on the cosmological 
model used and therefore these corrections will have to be worked separately for 
each model case studied. 

Studying the different types of scaling relations involving parameters derived from X-rays, we
can distinguish two types of scaling behavior: the mass - temperature relation is mostly
dependent on the scaling of the dark matter potentials and is therefore very close to the
gravitational scaling prediction. Most other relations involving gas density or gas mass
are affected by the non-constancy of the gas mass to total mass ratio. With the introduction
of modified scaling relations to take this hydrodynamical effects into account, we can
describe the currently available data sets within the given uncertainties.

Looking at the evolution of the scaling relations with redshift, we find an analogous
situation: the $M-T$ relation corresponds within the current uncertainties to the prediction
of the gravitational self-similar scenario. All other relations involving parameters
which depend on the gas density show deviations, which implies that the gas mass fraction is
not constant for given cluster mass with redshift. This is just the consequence of the
following effects. At higher redshifts clusters of given mass are more compact and the ICM has
to be squeezed into a deeper and narrower potential. Since in preheating models, which seem
to explain the data best, the gas starts out with an elevated entropy before cluster formation
is complete, the gas is less tightly squeezed into the earlier, narrower potentials than into 
the later wider potentials.  

Our modified scaling relation model does not describe all observational effects. As shown in 
Pratt et al. (2010), the entropy scaling depends on the radius at which the entropy is measured.
These results imply that the ICM depletion is larger in the center of groups and clusters than
in the outer parts. More data are required that extend out to large cluster radii (to $r_{500}$
and beyond) to substantiate this result.

The redshift evolution of the $L-T$ scaling relation, which is predicted to be $\propto E(z)$
in the simple self-similar scenario, is now found in recent studies to be much less positive or 
even negative (e.g. O'Hara et al. 2007, Reichert et al. 2011). This has the important consequence
that one will find less high redshift galaxy clusters in future X-ray and SZE surveys,
than predicted based on the simple scaling models (Reichert et al. 2011). More importantly,
a precise measurement of this evolution effect is crucial for using the future X-ray survey
data on galaxy clusters for the test of cosmological models.

As shown by the comparison of the evolution of the scaling relations compiled by Reichert
et al. (2011) and the simulations by Short et al. (2010), 
the study of the evolution of the scaling relations also provides
important insight into the astrophysics of the ICM. Currently the observational data strongly
favour a model with ealry preheating of the ICM. 

In the coming years both the observational data as well as the simulation results will experience
further strong improvements. Therefore we see the importance of this paper more in 
elucidating the way how the scaling relations should be analysed and applied, rather than already
providing the best parameterization of the results. 
 
\begin{acknowledgements} 

The paper is based on observations obtained with XMM-Newton, an ESA 
science mission with instruments and contributions directly funded by 
ESA Member States and the USA (NASA). The XMM-Newton project is 
supported by the Bundesministerium f\"ur Bildung und Forschung, 
Deutsches Zentrum f\"ur Luft und Raumfahrt (BMBF/DLR), the Max-Planck 
Society and the Haidenhain-Stiftung. H.B. acknowledges support from the DfG 
Transregio Programme TR33 and the Munich Excellence Cluster ''Structure and Evolution of the
Universe''. K.D. acknowledges support by the DfG Priority Programme 1177 and additional
support by the DfG Cluster of Excellence ''Structure and Evolution of the Universe''. 
We thank the anonymous referee for helpful comments.
 
\end{acknowledgements}

\end{document}